\documentclass[12pt]{article}
\usepackage{amstex,amsfonts}
\usepackage{graphicx}
\baselineskip =
1.3\normalbaselineskip

\begin{document}

\title{DISCOVERY OF PROTON DECAY
: A MUST FOR THEORY,  A CHALLENGE
FOR
EXPERIMENT\footnote{Invited talk
presented at the Int'l Workshop
on Next Generation Nucleon Decay
and Neutrino Detector, held
at Stony Brook, September, 23-25,
1999,to appear in the
Proceedings.}}
\author{Jogesh C. Pati\\[.8em]
{\normalsize \em Department of
Physics, University of Maryland,
College
Park}\\
{\normalsize \em MD 20742, USA.} }
\date{April 3, 2000}
\maketitle

\begin{abstract}
It is noted that, but for one missing piece -- proton decay -- the
evidence
 in support of grand unification is now strong. It
includes: (i) the
observed family-structure, (ii) the meeting of the gauge couplings,
(iii) neutrino-oscillations, (iv) the intricate pattern of the masses
and mixings of all fermions, including the neutrinos, and
 (v) the need for $B-L$ as a generator, to implement
baryogenesis. Taken together, these not only favor grand unification
but in fact select out a particular route to such unification, based
on the ideas of supersymmetry, SU(4)-color and left-right
symmetry. Thus they point to the relevance of an effective
string-unified G(224) or SO(10)-symmetry.

A concrete proposal is presented, within a
predictive SO(10)/G(224)-framework, that 
successfully describes the masses and mixings
of all fermions, including the neutrinos - with eight predictions, all in
agreement with observation. Within this
framework, a systematic study of proton decay is carried out, which
pays special attention to its dependence on the fermion masses,
including  the superheavy Majorana masses of the right-handed
neutrinos. The study shows that a conservative upper limit on the
proton lifetime is about (1/2 - 1)$\times10^{34}$ yrs, with
$\overline{\nu}K^{+}$ being the dominant decay mode, and as a
distinctive feature, $\mu^{+}K^{0}$ being prominent. This in turn strongly
suggests that an
improvement in the current sensitivity by a factor of five to
ten (compared to SuperK) ought to reveal proton decay. Otherwise some
promising and remarkably successful ideas on unification would suffer a
major setback.  \end{abstract}

\section{Introduction}

\noindent It has been recognized since the early 1970's that the price
one must pay to achieve a unification of quarks and leptons and
simultaneously a unity of the three gauge forces, commonly called
``grand unification'', is proton decay \cite{no1,no2,no3,no4}. This
important process, which would provide the window for viewing physics
at truly short distances ($<10^{-30}$ cm), is yet to be
seen. Nevertheless, as I will stress in this talk, there have appeared
over the years an impressive set of facts, including the meeting of
the gauge couplings and neutrino-oscillations, which not only favor
the hypothesis of grand unification, but in fact select out a
particular route to such unification, based on the ideas of
supersymmetry \cite{no5} and SU(4)-color \cite{no2}. These facts
together provide a clear signal that the discovery of proton decay
cannot be far behind.

To be specific, working within the framework of a unified theory
\cite{no6}, that incorporates the ideas mentioned above, I would argue
that an improvement in the current sensitivity for detecting proton
decay by a modest factor of five to ten should either produce real
events, or else the framework would be excluded. In this
sense, and as I will elaborate further, the discovery of proton decay
is now crucial to the survival of some elegant ideas on unification,
which are otherwise so successful. By the same token, proving or
disproving their prediction on proton decay poses a fresh 
challenge to experiment.

The pioneering efforts by several physicists \cite{no7} in the mid
1950's through the early 70's had provided a lower limit on the proton
lifetime of about $10^{26}$ yrs, independent of decay modes, and
$10^{29}$-$10^{30}$ yrs for the $e^{+}\pi^{0}$-mode. Subsequent
searches at the Kolar Goldfield and the NUSEX detectors in the early
80's \cite{no8} pushed this limit to about $10^{31}$ years in the
$e^{+}\pi^{0}$-mode. Following the suggestion of proton decay in the
context of grand unification, and thanks to the initiative of several
experimenters, two relatively large-size detectors - IMB and
Kamiokande - were built in the 80's, where dedicated searches for
proton decay were carried out with higher sensitivity. These detectors
helped to push the lower limit in the $e^{+}\pi^{0}$ channel to about
$10^{32}$ yrs. This in turn clearly disfavored the minimal {\it
non-supersymmetric} SU(5)-model of grand unification \cite{no3} - a
conclusion that was strengthened subsequently by the measurements of
the gauge couplings at LEP as well (see discussion later).

The searches for proton decay now continues with still greater
sensitivity at the largest detector so far - at SuperKamiokande,
completed in 1996. It is worth noting at this point that, although
these detectors have not revealed proton decay yet, they did bring
some major bonuses of monumental importance. These include : (a) the
detection of the neutrinos from the supernova 1987a, (b) confirmation
of the solar neutrino-deficit, and last but not least, (c) the
discovery of atmospheric neutrino-oscillation. The SuperK
water-Cerenkov detector with a fiducial volume of 22.5 kilotons
currently provides (with three years of running) a lower limit on the
inverse rate of proton decay of about $1.6\times10^{33}$ yrs for the
theoretically favored ($\overline{\nu}K^{+}$)-channel \cite{no9} and
of about $3.8\times10^{33}$ yrs for the ($e^{+}\pi^{0}$)-channel
\cite{no10}. It has the capability of improving these limits by a
factor of two to three in each case within the next decade, unless of
course it discovers real events for proton decay or strong candidate
events in the meantime. I will return to this point and its relevance
to theoretical expectations for proton decay in just a bit.

While proton decay is yet to be observed, it is worth stressing at
this point, that the hypothesis of grand unification, especially that
based on the ideas of SU(4)-color, left-right gauge symmetry, 
and supersymmetry, is now supported by several observations. As I
will explain in sections 2-5, these include :

{\bf (a)} {\bf The observed family structure :} The five scattered
multiplets of the standard model, belonging to a family, neatly become
parts of a whole ({\it a single multiplet}), with their weak
hypercharges precisely predicted by grand unification. Realization of
this feature calls for an extension of the standard model symmetry
G(213)$\,=\,$SU(2$)_{L}\times$U(1$)_{Y}\times$SU(3$)^{C}$ {\it
minimally} to the symmetry group
G(224)$\,=\,$SU(2$)_{L}\times$SU(2$)_{R}\times$SU(4$)^{C}$ \cite{no2},
which can be extended further into the simple group SO(10)
\cite{no11}, but not SU(5) \cite{no3}. The G(224) symmetry in turn
introduces some additional attractive features (see Sec. 2), including
especially the right-handed (RH) neutrinos ($\nu_{R}$'s) accompanying
the left-handed ones ($\nu_{L}$'s), and $B$-$L$ as a local
symmetry. As we will see, both of these features, which are special to
G(224), now seem to be needed on empirical grounds.

{\bf (b)} {\bf Meeting of the gauge couplings :} Such a meeting
is found to occur at a scale $M_{X}\approx2\times10^{16}$ GeV, when
the three gauge couplings are extrapolated from their values measured at
LEP
to higher energies, in the context of supersymmetry \cite{no12}. This
dramatic phenomenon supports the ideas of both grand unification and
supersymmetry. These in turn may well emerge from a string
theory \cite{no13} or M-theory \cite{no14} (see discussion in Sec. 3).

{\bf (c)} {\bf Mass of $\bf{\nu_{\tau}\sim1/20}$ eV :} Subject to the
well-motivated assumption of hierarchical neutrino masses, the recent
discovery of atmospheric neutrino-oscillation at SuperKamiokande
\cite{no15} suggests  a value for $m(\nu_{\tau})\sim1/20 eV$. It has
been argued (see e.g. Ref. \cite{no16}) that
a mass for $\nu_{\tau}$ of this magnitude points to the need for 
RH neutrinos, and that it goes extremely well with the hypothesis of a
supersymmetric unification, based on either a string-unified
G(224) symmetry or 
SO(10). The SUSY unification-scale as well as 
 SU(4)-color play crucial roles in making this
argument. 

{\bf (d)} {\bf Some intriguing features of fermion masses and
mixings :} These include : 
(i) the ``observed'' near equality of
the masses of the b-quark and the $\tau$-lepton at the
unification-scale ($m^{0}_{b}\approx m^{0}_{\tau}$);
(ii) the empirical Georgi-Jarsklog relations: $m^{0}_{s}\sim
m^{0}_{\mu}/3$ and $m^{0}_{d}\sim 3m^{0}_{e}$, and
(iii) the observed largeness of the
$\nu_{\mu}$-$\nu_{\tau}$  oscillation angle
($\sin^{2}2\theta^{\mbox{\scriptsize
osc}}_{\nu_{\mu}\nu_{\tau}}\geq0.83$) \cite{no15}, together with the
smallness of the corresponding quark mixing parameter
$V_{bc}(\approx0.04)$ \cite{no17}.
As shown in recent work
by Babu, Wilczek and me \cite{no6}, it turns out that these features
and more can be understood remarkably well (see discussion in Sec 5)
within an economical and predictive SO(10)-framework based on a
minimal Higgs system. The success of this framework is in large part
due simply to the group-structure of SO(10). For most purposes, that
of G(224) suffices.

{\bf (e)} {\bf Baryogenesis :} To implement baryogenesis \cite{no18}
successfully, in the presence of electroweak sphaleron effects
\cite{no19}, which wipe out out any baryon excess generated at high
temperatures in the ($B$-$L$)-conserving mode, it has become apparent
that one would need $B$-$L$ as a generator of the underlying symmetry,
whose spontaneous violation at high temperatures would yield, for
example, lepton asymmetry (leptogenesis). The latter in turn is
converted to baryon-excess at lower temperatures by electroweak
sphalerons. This mechanism, it turns out, yields even quantitatively
the right magnitude for baryon excess \cite{no20}. The need for
$B$-$L$, which  is a generator of SU(4)-color, again points to the
need for G(224) or SO(10) as an effective symmetry near the
unification-scale $M_{X}$.

The success of each of these five features (a)-(e) seems to be
non-trivial. Together they make a strong case for both
supersymmetric grand unification and simultaneously for the
G(224)/SO(10)-route to such unification, as being relevant to
nature. However, despite these successes, as long as proton decay
remains undiscovered, the hallmark of grand unification - that is {\it
quark-lepton transformability} - would remain unrevealed.

The relevant questions in this regard then are : What is the predicted
range for the lifetime of the proton - in particular an upper limit -
within the emperically favored route to unification mentioned above?
What are the expected dominant decay modes within this route? Are
these predictions compatible with current lower limits on proton
lifetime mentioned above, and if so, can they still be tested at the
existing or possible near-future detectors for proton decay?

Fortunately, we are in a much better position to answer these
questions now, compared to a few years ago, because meanwhile we have
learnt more about the nature of grand unification. As noted above (see
also Secs. 2 and 4), the neutrino masses and the meeting of the gauge
coupleings together seem to select the supersymmetric
G(224)/SO(10)-route to higher unification. The main purpose of my talk
here will therefore be to address the questions raised above, in the
context of this route. For the sake of comparison, however, I will
state the corresponding results for the case of supersymmetric SU(5)
as well.

My discussion will be based on a recent study of proton decay by Babu,
Wilczek and me \cite{no6},which, relative to previous ones,
has three distinctive features :

{\bf (a)} It systematically takes into account the link that exists
between proton decay and the masses and mixings of all fermions,
including the neutrinos. 

{\bf (b)} In particular, in adition to the contributions from the
so-called ``standard'' $d=5$ operators \cite{no22} (see Sec. 6), it
includes those from a {\it new} set of $d=5$ operators, related to the
Majorana masses of the RH neutrinos \cite{no21}. These latter are
found to be as important as the standard ones. 

{\bf (c)} The work also incorporates
 GUT-scale threshold effects,
 which arise because of mass-splittings
between the components of the SO(10)-multiplets, and lead to differences
 between the three gauge
couplings.

Each of these features
turn out to be {\it crucial} to gaining a reliable insight into the
nature of proton decay. Our study shows that the inverse decay rate
for the $\overline{\nu}K^{+}$-mode, which is dominant, is less than
about $7\times10^{33}$ yrs. This upper bound is obtained by making
generous allowance for uncertainties in the matrix elements and the
SUSY-spectrum. Typically, the lifetime should of course be less than
this bound. Furthermore, due to contributions from the new
operators, the $\mu^{+}K^{0}$-mode is found to be prominent, with a
branching ratio typically in the range of 10-50\%. By contrast,
minimal SUSY SU(5), for which the new operators are absent, would lead
to branching ratios $\leq10^{-3}$ for this mode. Thus our study of
proton decay , correlated with fermion masses, strongly suggests that
discovery of proton decay should be around the corner. In fact,one expects 
that
at least candidate events  should be observed in the near future
already at SuperK. However, allowing for the possibility that the
proton lifetime may well be closer to the upper bound stated above, a
next-generation detector providing a net gain in sensitivity in proton
decay-searches by a factor of 5-10, compared to SuperK, would
certainly be needed not just to produce proton-decay events, but also
to clearly distinguish them from the background. It would of course
also be essential  to study the branching ratios of certain
sub-dominant but crucial decay modes, such as the $\mu^{+}K^{0}$. The
importance of such improved sensitivity, in the light of the successes
of supersymmetric grand unification, is emphasized at the end.

\section{Advantages of the Symmetry G(224) as a Step to Higher Unification}

\noindent The standard model (SM) based on the gauge symmetry
G(213)$\,=\,$SU(2$)_{L}\times$U(1$)_{Y}\times$SU(3$)^{C}$ has turned
out to be extremely successful empirically. It has however been
recognized since the early 1970's that, judged on aesthetic merits, it
has some major shortcomings. For example, it puts members of a
family into five scattered multiplets, without providing a compelling
reason for doing so. It also does not provide a fundamental reason for
the quantization of electric charge. Nor does it explain the
co-existence of quarks and leptons, and that of the three gauge
forces, with their differing strengths. The idea of grand unification
was postulated precisely to remove these shortcomings. That in turn
calls for the existence of fundamentally new physics, far beyond that
of the standard model. As mentioned before, recent experimental
findings, including the meetings of the gauge couplings and
neutrino-oscillations, seem to go extremely well with this line of
thinking. 

To illustrate the advantage of an early suggestion in this regard,
consider the five standard model multiplets belonging to the
electron-family as shown : 
\begin{eqnarray}
\left(\begin{array}{ccc}{u_{r}}\,\,\,\,{u_{y}}\,\,\,\,{u_{b}}\\{d_{r}}\,\,\,\,{d_{y}}\,\,\,\,{d_{b}}\end{array}\right)^{\frac{1}{3}}_{L}\,;\,\,
\left(\begin{array}{ccc}{u_{r}}\,\,\,\,{u_{y}}\,\,\,\,{u_{b}}\end{array}\right)^{\frac{4}{3}}_{R}\,;\,\,
\left(\begin{array}{ccc}{d_{r}}\,\,\,\,{d_{y}}\,\,\,\,{d_{b}}\end{array}\right)^{-\,\frac{2}{3}}_{R}\,;\,\,
\left(\begin{array}{c}{\nu_{e}}\\{e^{-}}\end{array}\right)^{-\,1}_{L}\,;\,\,
\left(e^{-}\right)^{-\,2}_{R}\,.
\label{e1}
\end{eqnarray}
Here the superscripts denote the respective weak hypercharges $Y_{W}$
(where $Q_{em}=I_{3L}+Y_{W}/2$) and the subscripts L and R denote the
chiralities of the respective fields. If one asks : how one can put
these five multiplets into just one multiplet, the answer turns out to
be simple and unique. As mentioned in the introduction, the minimal
extension of the SM symmetry G(213) needed, to achieve this goal, is
given by the gauge symmetry \cite{no2} :
\begin{eqnarray}
\mbox{G(224)}\,=\,\mbox{SU(2})_{L}\times \mbox{SU(2})_{R}\times \mbox{SU(4})^{C}\,.
\label{e2}
\end{eqnarray}
Subject to left-right discrete symmetry ($L\leftrightarrow R$), which
is natural to G(224), all members of the electron family fall into the
neat pattern :
\begin{eqnarray}
F^{e}_{L,\,R}\,=\,\left[\begin{array}{cccc}{u_{r}}\,\,\,\,{u_{y}}\,\,\,\,{u_{b}}\,\,\,\,{\nu_{e}}\\{d_{r}}\,\,\,\,{d_{y}}\,\,\,\,{d_{b}}\,\,\,\,{{e}^{-}}\end{array}\right]_{L,\,R}
\label{e3}
\end{eqnarray}
The multiplets $F^{e}_{L}$ and $F^{e}_{R}$ are left-right conjugates
of each other and transform respectively as (2,1,4) and (1,2,4) of
G(224); likewise for the muon and the tau families. Note that the
symmetries SU(2$)_{L}$ and SU(2$)_{R}$ are just like the familiar
isospin symmetry, except that they operate on quarks and well as
leptons, and distinguish between left and right chiralities. The left
weak-isospin SU(2$)_{L}$ treats each column of $F^{e}_{L}$ as a
doublet; likewise SU(2$)_{R}$ for $F^{e}_{R}$; the symmetry
SU(4)-color treats each row of $F^{e}_{L}$ {\it and} $F^{e}_{R}$ as a
quartet, interpreting lepton number as the fourth color. Note also
that postulating either SU(4)-color or SU(2$)_{R}$ forces one to
introduce a right-handed neutrino ($\nu_{R}$) for each family as a
singlet of the SM symmetry. {\it This requires that there be sixteen
two-component fermions in each family, as opposed to fifteen for the
SM}. The symmetry G(224) introduces an elegant charge formula : 
\begin{eqnarray}
Q_{em}\,=\,I_{3L}\,+\,I_{3R}\,+\,\frac{B\,-\,L}{2}
\label{e4}
\end{eqnarray}
expressed in terms of familiar quantum numbers $I_{3L}$, $I_{3R}$ and
$B$-$L$, which applies to all forms of matter (including quarks and
leptons of all six flavors, gauge and Higgs bosons). Note that the
weak hypercharge given by $Y_{W}/2=I_{3R}\,+\,\frac{B\,-\,L}{2}$ is
now completely determined for all members of the family. The values of
$Y_{W}$ thus obtained precisely match the assignments shown in
Eq. (\ref{e1}). Quite clearly, the charges $I_{3L}$, $I_{3R}$ and
$B$-$L$, being generators respectively of SU(2$)_{L}$, SU(2$)_{R}$ and
SU(4$)^{c}$, are quantized; so also then is the electric charge
$Q_{em}$. 

In brief, the symmetry G(224) brings some attaractive features to
particle physics. These include :\\  (i) Organization of all 16
members of a family into one left-right self-conjugate multiplet;\\
(ii) Quantization of electric charge;\\  (iii) Quark-lepton
unification (through SU(4) color);\\  (iv) Conservation of parity at a
fundamental level \cite{no2,no23};\\  (v) Right-handed neutrinos
($\nu_{R}'s$) as a compelling feature; and\\  (vi) $B$-$L$ as a local
symmetry.\\  As mentioned in the introduction, the two distinguishing
features of G(224) - i.e. the existence of the RH neutrinos and
$B$-$L$ as a local symmetry - now seem to be needed on empirical
grounds.

Believing in a complete unification, one is led to view the G(224)
symmetry as part of a bigger symmetry, which itself may have its
origin in an underlying theory, such as string theory. In this
context, one might ask : Could the effective symmetry below the string
scale in four dimensions (see sec.3) be as small as just the SM
symmetry G(213), even though the latter may have its origin in a bigger
symmetry, which lives however only in higher
dimensions? I will argue in Sec. 4 that the data on neutrino masses
and the need for baryogenesis provide an answer to the contrary,
suggesting clearly that it is the effective symmetry in four
dimensions, below the string scale, which must {\it minimally} contain
either G(224) or a close relative
G(214)$\,=\,$SU(2$)_{L}\times$I$_{3R}\times$SU(4$)^{C}$. 

One may also ask : does the effective four dimensional symmetry  have
to be any bigger than G(224) near the string scale? In preparation for
an answer to this question, let us recall that the smallest simple
group that contains the SM symmetry G(213) is SU(5) \cite{no3}. It has
the virtue of demonstrating how the main ideas of grand unification,
including unification of the gauge couplings, can be
realized. However, SU(5) does not contain G(224) as a subgroup. As
such, it does not possess some of the advantages listed above. In
particular, it does not contain the RH neutrinos as a compelling
feature, and $B$-$L$ as a local symmetry. Furthermore, it splits
members of a family into two multiplets : $\overline{5}+10$. 

By contrast, the symmetry SO(10) has the merit, relative to SU(5),
that it contains G(224) as a subgroup, and thereby retains all the
advantages of G(224) listed above. (As a historical note, it is worth
mentioning that these advantages had
been motivated on aesthetic grounds  through the symmetry 
G(224) \cite{no2}, and {\it all} the ideas of higher unification were in
place
\cite{no1,no2,no3}, before it was noted that G(224)(isomorphic
to SO(4)$\times$SO(6)) embeds nicely into SO(10)
\cite{no11}). Now, {\it
SO(10) even preserves the 16-plet family-structure of G(224) without a
need for any extension}. By contrast, if one extends G(224) to the
still higher symmetry E$_{6}$ \cite{no24}, the advantages (i)-(vi) are
retained, but in this case, one must extend the family-structure from
a 16 to a 27-plet, by postulating additional fermions. In this sense,
there seems to be some advantage in having the effective symmetry
below the string scale to be minimally G(224) (or G(214)) and
maximally no more than SO(10). I will compare the relative advantage
of having either a string-derived G(224) or a string-SO(10), in the
next section. First, I discuss the implications of the data on
coupling unification.

\section{The Need for Supersymmetry : MSSM versus String Unifications}

It has been known for some time that the precision measurements of the
standard model coupling constants (in particular $\sin^{2}\theta_{W}$)
at LEP put severe constraints on the idea of grand unification. Owing
to these constraints, the non-supersymmetric minimal SU(5), and for
similar reasons, the one-step breaking minimal non-supersymmetric
SO(10)-model as well, are now excluded \cite{no25}. But the situation
changes radically if one assumes that the standard model is replaced
by the minimal supersymmetric standard model (MSSM), above a threshold
of about 1 TeV. In this case, the three gauge couplings are found to
meet \cite{no12}, at least approximately, provided $\alpha_{3}(m_{Z})$
is not too low (see Figs. in e.g. Refs. \cite{no23,no13}). Their scale
of meeting is given by
\begin{eqnarray}
M_{X}\approx 2\times10^{16}\,\mbox{GeV\,\,\,\,(MSSM or SUSY\,\,SU(5))}
\label{e5}
\end{eqnarray}

This dramatic meeting of the three gauge couplings, or equivalently
the agreement of the MSSM-based prediction of
$\sin^{2}\theta_{W}(m_{Z})_{\mbox{Th}}=0.2315\pm0.003$ \cite{no26}
with the observed value of
$\sin^{2}\theta_{W}(m_{Z})=0.23124\pm0.00017$ \cite{no17}, provides a
strong support for the ideas of both grand unification and
supersymmetry, as being relevant to physics at short distances. 

The most straightforward interpretation of the observed meeting of the
three couplings and of the scale $M_{X}$, is that a supersymmetric
grand unification symmetry (often called GUT symmetry), like SU(5) or
SO(10), breaks spontaneously at $M_{X}$ into the standard model
symmetry G(213).

In the context of string or M theory,
 which seems to be needed  to unify all the
forces of nature including gravity and also to obtain a good quantum
theory of gravity, an alternative interpretation is
however possible. This is because, even if the effective symmetry in
four dimensions emerging from a higher dimensional string theory is
non-simple, like G(224) or G(213), string theory can still ensure
familiar unification of the gauge couplings at the string scale. In
this case, however, one needs to account for the small mismatch
between the MSSM unification scale $M_{X}$ (given above), and the
 string unification scale, given by $M_{st}\approx
g_{st}\times5.2\times10^{17}$ GeV $\approx 3.6\times10^{17}$ GeV (Here we
have put
$\alpha_{st}=\alpha_{GUT}(\mbox{MSSM})\approx 0.04$)
\cite{no27}. Possible resolutions of this mismatch have been
proposed. These include : (i) utilizing the idea of {\it
string-duality} \cite{no28} which  allows a lowering of $M_{st}$
compared to the value shown above, or alternatively   (ii) the idea of
a {\it semi-perturbative} unification that assumes the existence of
two vector-like families, transforming as $(16+\overline{16})$, at the
TeV-scale. The latter
raises $\alpha_{GUT}$ to about 0.25-0.3 and simultaneously $M_{X}$, in
two loop, to about $(1/2-2)\times10^{17}$ GeV \cite{no29} (Other
mechanisms resolving the mismatch are reviewed in Refs. \cite{no30}
and \cite{no31}). In practice, a combination of the two mechanisms
mentioned above may well be relevant\footnote{I have in mind the
possibility of string-duality \cite{no28} lowering $M_{st}$ 
for the case of semi-perturbative unification (for which
$\alpha_{st}\approx$0.25, and thus, without the use of string-duality,
$M_{st}$ would be about $10^{18}$ GeV) to a value of about
(1-2)$\times10^{17}$ GeV (say), and semi-perturbative unification
\cite{no29}
raising the MSSM value of $M_{X}$ to about 5$\times10^{16}$ GeV$\approx$ 
$M_{st}$(1/2 to 1/4) (say). In
this case,
an intermediate symmetry like G(224) emerging at $M_{st}$ would be
effective only within the short gap between $M_{st}$ and $M_{X}$,
where it would break into G(213). Despite this short gap, one would
still have the benefits of SU(4)-color that are needed to
understand neutrino masses (see sec.4). At the same time, Since the
gap is so small, the couplings of G(224), unified at $M_{st}$ would
remain essentially so at $M_{X}$, so as to match with the ``observed''
coupling unification, of the type suggested in Ref. \cite{no29}.}.

While the mismatch can thus quite plausibly be removed for a non-GUT
string-derived symmetry like G(224) or G(213), a GUT symmetry like
SU(5) or SO(10) would have an advantage in this regard because it
would keep the gauge couplings together between $M_{st}$ and $M_{X}$
(even if $M_{X}\sim M_{st}/20$), and thus  not even encounter the
problem of a mismatch between the two scales. A supersymmetric
GUT-solution (like SU(5) or SO(10)), however, has a possible
disadvantage as well, because it needs certain color triplets to
become superheavy by the so-called double-triplet splitting mechanism
(see Sec. 6 and Appendix), in order to avoid the problem of rapid
proton decay. However, no such mechanism has emerged yet, in string
theory, for the GUT-like solutions \cite{no32}.

Non-GUT string solutions, based on symmetries like G(224) or G(2113)
for example, have a distinct advantage in this regard, in that the
dangerous color triplets, which would induce rapid proton decay, are
often naturally projected out for such solutions
\cite{no33,no34}. Furthermore, the non-GUT solutions invariably
possess new ``flavor'' gauge symmetries, which distinguish between
families. These symmetries are immensely helpful in explaining
qualitatively the observed fermion mass-hierarchy (see
e.g. Ref. \cite{no34}) and resolving the so-called naturalness
problems of supersymmetry such as those pertaining to the issues of
squark-degeneracy \cite{no35}, CP violation \cite{no36} and quantum
gravity-induced rapid proton decay \cite{no37}.

Weighing the advantages and possible disadvantages of both, it seems
hard at present to make a priori a clear choice between a GUT versus
a non-GUT string-solution. As expressed elsewhere \cite{no31}, it
therefore seems prudent to keep both options open and pursue their
phenomenological consequences. Given the advantages of G(224) or
SO(10) in the light of the neutrino masses (see Secs. 2 and 4), I will
thus proceed by assuming that either a suitable G(224)-solution with a
mechanism of the sort mentioned above, or a realistic SO(10)-solution
with the needed doublet-triplet mechanism, will emerge from string
theory. We will see that with this broad assumption an economical and
predictive framework emerges, which  successfully accounts for a host of
observed phenomena, and makes some crucial testable predictions. 
Fortunately, it will turn out that there are many similarities between the
predictions of a string-unified G(224) and
SO(10),
not only for the neutrino and the charged fermion masses, but also for
proton decay. I next discuss the implications of the mass of
$\nu_{\tau}$ suggested by the SuperK data.

\section{Mass of $\nu_\tau$: Evidence In Favor of the G(224) Route}

One can obtain an estimate for the mass of $\nu_{L}^{\tau}$ in the
context of G(224) or SO(10) by using the following three steps
(see e.g.Ref.\cite{no16}): 
 
(i) Assume that B$-$L and $I_{3R}$, contained in a string-derived
G(224) or SO(10), break near the unification-scale:  
\begin{eqnarray} 
M_X\sim2\times10^{16}\,\mbox {GeV}\,, 
\label{e6}
\end{eqnarray} 
through VEVs of Higgs multiplets of the type suggested by
string-solutions - i.e. $\langle(1,2,4)_H\rangle$ for
G(224) or $\langle\overline{16}_{H}\rangle$ for  SO(10), as opposed to
$126_H$ \cite{no38}.  In the process, the RH neutrinos  ($\nu_{R}^{i}$),
which are
singlets of the standard model, can and  generically will acquire
superheavy Majorana masses of the type
$M_{R}^{ij}\,\nu_R^{iT}\,C^{-1}\,\nu_R^{j}$, by utilizing the VEV of
$\langle{\overline{16}}_{H}\rangle$ and effective  couplings of the
form: 
\begin{eqnarray} 
{\cal
L}_M\,(SO(10))\,=\,f_{ij}\,\,16_{i}\cdot16_{j}\,\,\overline{16}_{H}\,\cdot\overline{16}_{H}/M
+ h.c. 
\label{e7}
\end{eqnarray} 
 
A similar expression holds for G(224). Here $i,j=1,2,3$, correspond
respectively to $e,\,\mu$ and $\tau$ families.  Such gauge-invariant
non-renormalizable couplings might be expected to be induced by
Planck-scale  physics, involving quantum gravity or stringy effects
and/or tree-level exchange of superheavy  states, such as those in the
string tower.  With $f_{ij}$ (at least   the largest among them) being
of order unity, we would thus expect M to lie   between
$M_{Planck}\approx2\times10^{18}$ GeV and
$M_{string}\approx4\times10^{17}$ GeV. Ignoring for the present
off-diagonal mixings (for simplicity), one thus obtains 
\footnote{The effects of neutrino-mixing and of possible choice of 
$M=M_{string}\approx4\times 10^{17}$ GeV (instead of $M=M_{Planck}$) 
on $M_{3R}$ are considered in Ref. \cite{no6}.}:  
\begin{eqnarray} 
M_{3R}\,\approx\,\frac{f_{33}\langle\overline{16}_{H}\rangle^{2}}{M} 
\,\approx\,f_{33}\,(2\times10^{14}\,\mbox{GeV})\,\eta^{2}\,(M_{Planck}/M) 
\label{e8}
\end{eqnarray} 
 
This is the Majorana mass of the RH tau neturino. Guided by the 
value of $M_{X}$, we have substituted 
$\langle\overline{16}_{H}\rangle=(2\times10^{16}\,\mbox{GeV})\,\eta$ 
,with $\eta\approx1/2$ to 2(say).  
 
(ii) Now using SU(4)-color and the
Higgs  multiplet $(2,2,1)_{H}$ of G(224) or equivalently $10_{H}$ of
SO(10), one  obtains the relation $m_{\tau}(M_{X}) = m_{b}(M_{X})$,
which is known to be  successful. Thus, there is a good reason to
believe that the third family gets its masses primarily from the
$10_{H}$ or equivalently $(2,2,1)_{H}$ (see sec.5). In turn, this implies: 
\begin{eqnarray} 
m(\nu^{\tau}_{Dirac})\,\approx\,m_{top}(M_{X})\,\approx\,(100\,\mbox{-}\,120)\,\mbox{GeV} 
\label{e9}
\end{eqnarray} 
Note that this relationship between the Dirac mass of  the tau-neutrino 
and the top-mass is special to SU(4)-color.  It does not emerge in 
SU(5). 
 
(iii) Given the superheavy Majorana masses of the RH neutrinos as well 
as the Dirac masses as above, the see-saw mechanism \cite{no39} 
yields naturally light masses for the LH neutrinos.  For $\nu_{L}^{\tau}$ 
(ignoring mixing), one thus obtains, using Eqs. (\ref{e8}) and (\ref{e9}),  
\begin{eqnarray} 
m(\nu^{\tau}_{L})\,\approx\,\frac{m(\nu^{\tau}_{Dirac})^2}{M_{3R}}\,\approx\,[(1/20)\,\mbox{eV}\,(1\,\mbox{-}\,1.44)/f_{33}\,\eta^{2}]\,(M/M_{Planck}) 
\label{e10}
\end{eqnarray} 
 
Now,assuming the hierarchical pattern $m(\nu^{e}_{L})\ll
m(\nu^{\mu}_{L})\ll m(\nu^{\tau}_{L})$, which is suggested by the see-saw
mechanism,and further that the SuperK observation represents
$\nu_{L}^{\mu}-\nu_{L}^{\tau}$ (rather than $\nu_{L}^{\mu}-\nu_{X}$)
oscillation, the observed  $\delta
m^2\approx1/2(10^{-2}\,\mbox{-}\,10^{-3})\,$eV$^{2}$ corresponds  to
$m(\nu_{L}^{\tau})\approx$ (1/15 - 1/40) eV. It seems  {\it
truly remarkable} that the expected magnitude of $m(\nu_{L}^{\tau})$,
given by Eq.(\ref{e10}), is just about what is suggested by the SuperK
data, if
$f_{33}\,\eta^2(M_{Planck}/M)\approx$ 1.3 to 1/2.  Such a range for
 $f_{33}\,\eta^2(M_{Planck}/M)$ seems most plausible and
natural (see discussion in Ref. \cite{no16}). Note that
the estimate (\ref{e10}) crucially depends upon the supersymmetric
unification scale, which provides a value for $M_{3R}$,
as well as on SU(4)-color that yields $m(\nu^{\tau}_{Dirac})$.{\it The 
agreement between the expected and the SuperK result thus clearly
suggests that the effective symmetry below the string-scale should
contain SU(4)-color. Thus, minimally it should be either G(214) or
G(224), and maximally as big as SO(10), if not E$_{6}$}. 
 
By contrast, if SU(5) is regarded as either a fundamental symmetry or as
the effective symmetry below the string scale, there would be no
compelling reason based on symmetry alone, to introduce a $\nu_{R}$,
because it is a singlet of SU(5).  Second, even if one did introduce
$\nu^{i}_{R}$ by hand, their Dirac masses, arising from the coupling
$h^{i}\,\overline{5}_{i}\langle 5_H\rangle\nu^{i}_{R}$, would be
unrelated to the up-flavor  masses and thus rather arbitrary (contrast
with Eq. (\ref{e9})).  So also  would be the Majorana masses of the
$\nu^{i}_{R}$'s, which are  SU(5)-invariant, and thus can be even of
order string scale . This would give
$m(\nu^{\tau}_{L})$ in gross conflict  with the observed value. 

Before passing to the next section, it is worthnoting that the mass of
$\nu_{\tau}$ suggested by SuperK, as well as the observed value of
$\sin^{2}\theta_{W}$ (see Sec.3), provide valuable insight into the nature
of GUT symmetry breaking. They both favor the case of a single-step
breaking (SSB) of SO(10) or a string-unified G(224) symmetry at a scale
of order $M_{X}$, into the standard model symmetry G(213), as opposed to
that of a multi-step breaking (MSB). The latter would correspond, for
example, to SO(10) (or G(224)) breaking at a scale $M_{1}$ into G(2213),
 which in turn breaks at a scale $M_{2}<< M_{1}$ into G(213).
 One reason why the case of single-step  breaking is favored over that of
multi-step breaking is that the latter can accomodate but not really
predict $\sin^{2}\theta_{W}$, where as the former predicts the same
successfully. Furthermore, since the Majorana mass of $\nu^{\tau}_{R}$
arises arises only after $B-L$ and $I_{3R}$ break, it would be given, for
the case of MSB, by $M_{3R}\sim f_{33}(M_{2}^{2}/M)$, where $M\sim M_{st}$
(say). If $M_{2}\ll M_{X}\sim 2\times10^{16}$ GeV, and M $>M_{X}$, one
would obtain too low a value ($<< 10^{14}$ GeV) for $M_{3R}$ (compare with
Eq.(8)),and thereby too large a value for $m(\nu^{\tau}_{L})$, compared to
that suggested by SuperK. By contrast, the case of SSB yields the right
magnitude for $m(\nu_{\tau})$ (see Eq. (10)).

Thus the success of the
result on $m(\nu_{\tau})$ discussed above not only favors the symmetry
G(224) or SO(10), but also clearly suggests that $B-L$ and $I_{3R}$ break
near the conventional GUT scale $M_{X}\sim 2\times 10^{16}$ GeV, rather
than at an intermediate scale $<< M_{X}$. In other words, 
the
observed
values of both $\sin^{2}\theta_{W}$ and $m(\nu_{\tau})$ favor
only {\it the
simplest pattern of symmetry-breaking}, for which SO(10) or a
string-derived G(224) symmetry breaks in one step to
the
standard model symmetry, rather than in multiple steps. It is of course
only this simple
pattern of symmetry breaking that would be rather restrictive as regards
its predictions for proton decay (to be
dicussed in Sec.6). I next dicuss the problem of understanding the masses
and mixings of all fermions.

\section{Understanding Fermion Masses and Neutrino Oscillations in SO(10)}

Understanding the masses and mixings of all quarks and charged
leptons, in conjunction with those of the neutrinos, is a goal worth
achieving by itself. It also turns out to be essential for the study
of proton decay. I therefore present first a recent attempt in this
direction, which seems most promising \cite{no6}. A few guidelines
would prove to be helpful in this regard. The first of these is
motivated by the desire for economy and the rest by data. 

{\bf 1) Hierarchy Through Off-diagonal Mixings :} Recall earlier
attempts \cite{no40} that attribute hierarchical masses of the first
two families to matrices of the form :
\begin{eqnarray}
M\,=\,\left(\begin{array}{cc}{0}\,\,\,\,\,\,\,\,\,\,\,\,{\epsilon}\\{\epsilon}\,\,\,\,\,\,\,\,\,\,\,\,{1}\end{array}\right)\,m^{(0)}_{s}\,,
\label{e11}
\end{eqnarray}
for the $(d,s)$ quarks, and likewise for the $(u,c)$ quarks. Here
$\epsilon\sim1/10$. The hierarchical patterns in Eq. (\ref{e11}) can
be ensured by imposing a suitable flavor symmetry which distinguishes
between the two families (that in turn may have its origin in string
theory (see e.g. Ref \cite{no34}). Such a pattern has the virtues that
(a) it yields a hierarchy that is much larger than the input
parameter $\epsilon$ : $(m_{d}/m_{s})\approx\epsilon^{2}\ll\epsilon$,
and (b) it leads to an expression for the cabibbo
angle :
\begin{eqnarray}
\theta_{c}\approx\bigg|\sqrt{\frac{m_{d}}{m_{s}}}\,-\,e^{i\phi}\,\sqrt{\frac{m_{u}}{m_{c}}}\,\bigg|\,,
\label{e12}
\end{eqnarray}
which is rather successful. Using $\sqrt{m_{d}/m_{s}}\approx 0.22$ and
$\sqrt{m_{u}/m_{c}}\approx
0.06$, we see that Eq. (\ref{e12}) works to within about $25\%$ for
any value of the phase $\phi$. Note that the square root formula (like
$\sqrt{m_{d}/m_{s}}$) for the relevant mixing angle arises because of
the symmetric form of $M$ in Eq. (\ref{e11}), which in turn is ensured
if the contributing Higgs is a 10 of SO(10). A generalization of the
pattern in Eq. (\ref{e11}) would suggest that the first two families
(i.e. the $e$ and the $\mu$) receive masses primarily through
their mixing with the third family $(\tau)$, with $(1,3)$ and $(1,2)$
elements being smaller than the $(2,3)$; while  $(2,3)$ is smaller than
the
$(3,3)$. We will follow this guideline, except for the modification
noted below.

{\bf 2) The Need for an Antisymmetric Component :} Although the
symmetric hierarchical matrix in Eq. (\ref{e11}) works well for the
first two families, a matrix of the same form fails altogether to
reproduce $V_{cb}$, for which it yields :
\begin{eqnarray}
V_{cb}\approx\bigg|\sqrt{\frac{m_{s}}{m_{b}}}\,-\,e^{i\chi}\,\sqrt{\frac{m_{c}}{m_{t}}}\,\bigg|\,.
\label{e13}
\end{eqnarray}
Given that $\sqrt{m_{s}/m_{b}}\approx 0.17$ and
$\sqrt{m_{c}/m_{t}}\approx 0.0.06$, we see that Eq. (\ref{e13}) would
yield $V_{cb}$ varying between 0.11 and 0.23, depending upon the phase
$\chi$. This is too big, compared to the observed value of
$V_{cb}\approx0.04\pm0.003$, by at least a factor of 3. We interpret
this failure as a {\it clue} to the presence of an antisymmetric
component in $M$, together with symmetrical ones
(thus $m_{ij}\neq m_{ji}$), which would modify the relevant mixing angle
to
$\displaystyle{\sqrt{\frac{m_{i}}{m_{j}}}\sqrt{\frac{m_{ij}}{m_{ji}}}}$,
where $m_{i}$ and $m_{j}$ denote the respective eigenvalues. 

{\bf 3) The Need for a Contribution Proportional to $B$-$L$ :} The
success of the relations $m^{0}_{b}\approx m^{0}_{\tau}$, and
$m^{0}_{t}\approx m(\nu_{\tau})^{0}_{Dirac}$ (see Sec. 4), suggests
that the members of the third family get their masses primarily from
the VEV of a SU(4)-color singlet Higgs field that is independent of
$B$-$L$. This is in fact ensured if the Higgs is a 10 of
SO(10). However, the empirical observations of
$m^{0}_{s}\sim m^{0}_{\mu}/3$ and $m^{0}_{d}\sim 3m^{0}_{e}$ \cite{no41} 
clearly
call for a contribution proportional to $B$-$L$ as well. Further, one
can in fact argue that the suppression of $V_{bc}$ (in the
quark-sector) together with an enhancement of
$\theta^{osc}_{\nu_{\mu}\,\nu_{\tau}}$ (in the lepton sector) calls
for a contribution that is not only proportional to $B$-$L$ but is
also antisymmetric in the family space (as suggested above in item 
(\ref{e2})). We note below how both of these requirements can be met,
rather easily,
 in
SO(10), even for a minimal Higgs system.

{\bf 4) Up-Down Asymmetry :} Finally, the up and
the down-sector mass matrices must not be proportional to each other, as
otherwise the CKM angles would all vanish. 

Following Ref. \cite{no6}, I now present a simple and predictive
mass-matrix, based on SO(10), that satisfies {\it all three}
requirements, (\ref{e2}), (\ref{e3}) and (\ref{e4}). The
interesting point is that one can obtain such a mass-matrix for the
fermions by utilizing only the minimal Higgs system, that is needed
anyway to break the gauge symmetry SO(10). It consists of the set :
\begin{eqnarray}
H_{minimal}\,=\,\{45_{H},\,16_{H},\,\overline{16}_{H},\,10_{H}\}\,.
\label{e14}
\end{eqnarray}
Of these, the VEV of $\langle45_{H}\rangle\sim M_{X}$ breaks SO(10)
into G(2213), and those of
$\langle16_{H}\rangle=\langle\overline{16}_{H}\rangle\sim M_{X}$ break
G(2213) to G(213), at the unification-scale $M_{X}$. Now G(213) breaks
at the electroweak scale by the VEV of $\langle10_{H}\rangle$ to
U$(1)_{em}\times$ SU$(3)^{c}$. 

One might have introduced large-dimensional tensorial multiplets of
SO(10) like 12$6_{H}$ and 12$0_{H}$, both of which possess 
cubic level Yukawa couplings with the fermions. In particular, the
coupling $16_{i}16_{j}(120_{H})$ would give the desired
family-antisymmetric as well as ($B$-$L$)-dependent contribution. We do
not however introduce these multiplets in part because they do not
seem to arise in string solutions \cite{no38}, and in part also
because mass-splittings within such large-dimensional multiplets tend
to give excessive threshold corrections to $\alpha_{3}(m_{z})$
(typically exceeding 20\%), rendering observed coupling
unification fortuitous. By contrast, the multiplets in the minimal set
(shown above) do arise in string solutions leading to
SO(10). Furthermore, the 
threshold corrections for the minimal set are found to be
 naturally small, and even to have the right sign,
to go with the observed coupling unification \cite{no6}.

The question is : does the minimal set meet all the requirements listed
above? Now $10_{H}$ (even several 10`s) can not 
meet the requirements of antisymmetry and
$(B$-$L)$-dependence. Furthermore, a single $10_{H}$ cannot generate
CKM-mixings. This impasse disappears,however, as soon as one allows
 for not only
cubic, but also effective non-renormalizable quartic couplings of the
minimal set of Higgs fields with the fermions. These latter couplings
could of course well arise through exchanges of superheavy states
(e.g. those in
the string tower) involving renormalizable couplings, and/or through
quantum gravity. 

Allowing for such cubic and quartic couplings and adopting the
guideline (\ref{e1}) of hierarchical Yukawa couplings, as well as that
of economy, we are led to suggest the following effective lagrangian
for generating Dirac masses and mixings of the three families
\cite{no6} (for a related but different pattern, involving a
non-minimal Higgs system, see Ref \cite{no42}).
\begin{eqnarray}
{\bf {\cal L}_{Yuk}}\,=\,h_{33}\,{\bf
16_{3}\,16_{3}\,10_{H}}\,+\,[\,h_{23}\,{\bf
16_{2}\,16_{3}\,10_{H}}\,+\,a_{23}\,{\bf
16_{2}\,16_{3}\,10_{H}\,45_{H}}/M\,\nonumber\\&&{\hspace{-12cm}}+\,g_{23}\,{\bf
16_{2}\,16_{3}\,16_{H}\,16_{H}}/M]\,+\,\{a_{12}\,{\bf
16_{1}\,16_{2}\,10_{H}\,45_{H}}/M\,\nonumber\\&&{\hspace{-10.5cm}}+\,g_{12}\,{\bf
16_{1}\,16_{2}\,16_{H}\,16_{H}}/M\}\,.
\label{e15}
\end{eqnarray}
Here, $M$ could plausibly be of order string scale. Note that a mass
matrix having essentially the form of Eq. (\ref{e11})  results if the
first term $h_{33}\langle10_{H}\rangle$ is dominant. This ensures
$m^{0}_{b}\approx m^{0}_{\tau}$ and $m^{0}_{t}\approx
m(\nu_{Dirac})^{0}$. Following the assumption of progressive hierarchy
(equivalently appropriate flavor symmetries\footnote{Although no
explicit string solution with the hierarchy in $h_{ij}$ mentioned
above, together with the $a_{ij}$ and $g_{ij}$ couplings of
Eq. (\ref{e15}), exists as yet, flavor symmetries of the type alluded
to, as well as SM singlets carrying flavor-charges and acquiring VEVs
of order $M_{X}$ that can lead to effective hierarchical couplings, do
emerge in string solutions. And, there exist solutions with top Yukawa
coupling being leading (see e.g. Refs. \cite{no34} and \cite{no33}).}),
we presume that $h_{23}\sim h_{33}/10$, while $h_{22}$ and $h_{11}$,
which are set to be zeros, are progressively much smaller than
$h_{23}$ (see discussion in Ref. \cite{no31}). Since
$\langle45_{H}\rangle\sim\langle16_{H}\rangle\sim M_{X}$, while $M\sim
M_{st}\sim10M_{X}$, the terms $a_{23}\langle45_{H}\rangle/M$ and
$g_{23}\langle16_{H}\rangle/M$ can quite plausibly be of order
$h_{33}/10$, if $a_{23}\sim g_{23}\sim h_{33}$. By the assumption of
hierarchy, we presume that $a_{12}\ll a_{23}$, and $g_{12}\ll g_{23}$

It is interesting to observe the symmetry properties of the $a_{23}$
and $g_{23}$-terms.  Although $10_{H}\times45_{H}=10+120+320$, given
that $\langle45_{H}\rangle$ is along $B$-$L$, which is needed to
implement doublet-triplet  splitting (see Appendix), only 120 in the
decomposition contributes to the  mass-matrices.  This contribution
is, however, antisymmetric in the  family-index and, at the same time,
proportional to $B$-$L$. {\it Thus the  $a_{23}$ term fulfills the
requirements of both antisymmetry and ($B$-$L$)-dependence,
simultaneously\footnote{The analog of $10_{H}\cdot45_{H}$ for the case
of G(224) would be
$\chi_{H}\equiv(2,2,1)_{H}\cdot(1,1,15)_{H}$. Although in general, the
coupling of $\chi_{H}$ to the fermions need not be antisymmetric, for
a string-derived G(224), the multiplet (1,1,15$)_{H}$ is most likely
to arise from an underlying 45 of SO(10) (rather than 210); in this
case, the couplings of $\chi_{H}$ must be antisymmetric like that of
$10_{H}\cdot45_{H}$.}}. With only $h_{ij}$ and
$a_{ij}$-terms,
however, the up and down quark  mass-matrices will be proportional to
each other, which would yield $V_{CKM} =1$.  This is remedied by the
$g_{ij}$ coupling. Because, the $16_{H}$ can have a VEV not only along
its SM singlet component (transforming as
$\tilde{\overline{\nu}}_{R}$) which is of GUT-scale, but also along
its electroweak doublet component -- call it $16_{d}$ -- of the
electroweak  scale. The latter can arise by the  the mixing of
$16_{d}$ with the corresponding doublet (call it $10_{d}$) in the
$10_{H}$.  The MSSM doublet $H_{d}$, which is light, is then a mixture
of $10_{d}$ and $16_{d}$, while the orthogonal combination is
superheavy (see Appendix). Since $\langle16_{d}\rangle$ contributes
only to the down-flavor mass matrices, but not to the up-flavor, the
$g_{23}$ and $g_{12}$ couplings  generate non-trivial
CKM-mixings. We thus see that the minimal  Higgs system satisfies
apriori all the qualitative requirements (2)-(4), including the
condition of $V_{CKM}\neq1$. I now discuss that this  system works
well even quantitatively.

With these six effective Yukawa couplings, the Dirac mass matrices  of
quarks and leptons of the three families  at the unification scale
take the form :
\begin{eqnarray} 
U\,=\,\left(\begin{array}{ccc}{0} & {\epsilon'} & {0} \\
{-\,\epsilon'} & {0}  & {\epsilon\,+\,\sigma} \\ {0} &
{-\,\epsilon\,+\,\sigma} & {1}\end{array}\right)\,m_U,\,\,\,\,
D\,=\,\left(\begin{array}{ccc}{0} & {\epsilon'\,+\,\eta'} & {0} \\
{-\,\epsilon'\,+\,\eta'} & {0} & {\epsilon\,+\,\eta} \\ {0} &
{-\,\epsilon\,+\,\eta} & {1}\end{array}\right)\,m_D,  \nonumber \\[1.5em]
N=\left(\begin{array}{ccc}{0} & {-\,3\epsilon'} & {0} \\ {3\epsilon'}
& {0} & {-\,3\epsilon\,+\,\sigma} \\ {0} & {3\epsilon\,+\,\sigma} &
{1}\end{array}\right)\,m_U,\,  L=\left(\begin{array}{ccc}{0} &
{-\,3\epsilon'\,+\,\eta'} & {0} \\ {3\epsilon'\,+\,\eta'} & {0} &
{-\,3\epsilon\,+\,\eta} \\ {0} & {3\epsilon\,+\,\eta} &
{1}\end{array}\right)\,m_D. 
\label{e16}
\end{eqnarray} 
Here the matrices are multiplied by  left-handed fermion fields  from
the left and by anti--fermion fields from  the right.  $(U,D)$ stand
for the mass matrices of up and  down quarks, while $(N,L)$ are the
Dirac mass matrices  of the neutrinos and the charged leptons. The
entries $1,\epsilon$,and $\sigma$ arise respectively from the
$h_{33},a_{23}$ and $h_{23}$ terms in Eq. (\ref{e15}), while $\eta$
entering into $D$ and $L$ receives contributions from both  $g_{23}$
and $h_{23}$; thus $\eta\neq\sigma$. Similarly $\eta'$ and $\epsilon'$
arise from $g_{12}$ and $a_{12}$ terms respectively. Note the
quark-lepton correlations between $U$ and $N$ as well as $D$ and $L$,
and the up-down correlations between $U$ and $D$ as well as $N$ and
$L$. These correlations arise because of the symmetry property of
G(224). The relative factor of $-3$ between  quarks and leptons
involving the $\epsilon$ entry reflects the fact  that $\langle{\bf
45_{H}}\rangle\propto(B-L)$, while the  antisymmetry in this entry
arises from the group structure of SO(10),  as explained
above$^4$. As we will see,
this $\epsilon$-entry helps to account for (a) the differences between
$m_{s}$ and $m_{\mu}$, (b) that between $m_{d}$ and $m_{e}$, and also,
(c) the suppression of $V_{cb}$ together with the enhancement of the
$\nu_{\mu}$-$\nu_{\tau}$ oscillation angle. 

The mass matrices in Eq. (\ref{e16}) contain 7 parameters\footnote{Of
these,  $m_{U}^{0}\approx m_{t}^{0}$ can in fact be estimated to
within $20\%$  accuracy by either using the argument of radiative
electroweak symmetry  breaking, or some promising string solutions
(see e.g. Ref. \cite{no34}).} : $\epsilon$, $\sigma$, $\eta$,
$m_{D}=h_{33}\,\langle10_{d}\rangle$,
$m_{U}=h_{33}\,\langle10_{U}\rangle$, $\eta'$ and $\epsilon'$. These
may be determined by using, for  example, the following input values:
$m_{t}^{phys}=174$ GeV, $m_{c}(m_{c})=1.37$ GeV, $m_{s}(1$
GeV$)=110$-$116$ MeV \cite{no43}, $m_{u}(1$ GeV) $\approx6$ MeV and
the observed masses of $e$, $\mu$ and $\tau$, which lead to (see
Ref. \cite{no6}, for details) :
\begin{eqnarray} 
\sigma\,\simeq\,\,0.110\,,\,\,\eta\simeq\,0.151\,,\,\,\epsilon\,\simeq\,-\,0.095\,,\,\,|\eta'|\approx4.4\times10^{-3}\,\,\,\mbox{and}\,\,\,\epsilon'\approx2\times10^{-4}\nonumber  
\end{eqnarray}
\begin{eqnarray}
m_{U}\,\simeq\,m_{t}(M_{U})\,\simeq\,(100\mbox{-}120)\,\mbox{GeV}\,,\,\,m_{D}\,\simeq\,m_{b}(M_{U})\,\simeq\,1.5\,\mbox{GeV}\,.
\label{e17} 
\end{eqnarray} 
We have assumed for simplicity that the parameters are real, because a
good fitting suggests that the relative phases of at least $\sigma$,
$\eta$ and $\epsilon$ are small ($<10^{\circ}$ say).Such fitting also
fixes
their relative signs. Note that in accord with our general
expectations  discussed above, each of the parameters $\sigma$, $\eta$
and $\epsilon$ are found to be of order 1/10,  as opposed to being
\footnote{This is one characteristic difference  between our work and
that of Ref. \cite{no38}, where the (2,3)-element is even  bigger than
the (3,3).} $O(1)$ or $O(10^{-2})$, compared to the leading
(3,3)-element in Eq. (\ref{e16}). Having determined these parameters,
we are led to a total of five predictions involving only the quarks
(those for the leptons are listed separately) :
\begin{eqnarray} 
 m^{0}_{b}\,\approx\,m^{0}_{\tau}(1\,-\,8\epsilon^{2})\,;\,\,\,\,
\mbox{thus}\,\,\,\,m_{b}(m_{b})\,\simeq\,(4.6\mbox{-}4.9)\,\mbox{GeV}
\label{e18}
\end{eqnarray} 
\vspace*{-2em}
\begin{eqnarray} 
|V_{cb}|\,\simeq\,|\sigma\,-\,\eta|\,\approx\,
\left|\sqrt{m_{s}/m_{b}}\left|\frac{\eta\,+
\,\epsilon}{\eta\,-\,\epsilon}\right|^{1/2}\,-
\,\sqrt{m_{c}/m_{t}}\,\left|\frac{\sigma\,+\,\epsilon}{\sigma\,-
\,\epsilon}\right|^{1/2}\right|\,\simeq\,0.045 
\label{e19} 
\end{eqnarray} 
\vspace*{-2em}
\begin{eqnarray} 
m_{d}\,(1 \mbox{GeV})\,\simeq\,8\,\mbox{MeV}
\label{e20} 
\end{eqnarray} 
\vspace*{-2em}
\begin{eqnarray} 
\theta_{C}\,\simeq\,\left|\sqrt{m_{d}/m_{s}}\,-
 \,e^{i\phi}\sqrt{m_{u}/m_{c}}\right|
\label{e21} 
\end{eqnarray} 
\vspace*{-2em}
\begin{eqnarray} 
|V_{ub}/V_{cb}|\,\simeq\,\sqrt{m_{u}/m_{c}}\,\simeq\,0.07\,.
\label{e22} 
\end{eqnarray} 
In making these predictions, we have extrapolated the GUT-scale values
down to low energies using $\alpha_{3}(m_{Z})=0.118$, a SUSY threshold
of 500 GeV and $\tan\beta=5$. The results depend  weakly on these
choices, assuming  $\tan\beta\approx2$-30. Further, the Dirac masses and
mixings of the neutrinos and the mixings of the  charged  leptons also
get determined. We obtain :
\begin{eqnarray} 
m_{\nu_{\tau}}^{D}(M_{U})\,\approx\,100\mbox{-}120\,
\mbox{GeV};\,\,m_{\nu_{\mu}}^{D}(M_{U})\,\simeq\,8\,\mbox{GeV},
\label{e23} 
\end{eqnarray} 
\vspace*{-2em}
\begin{eqnarray} 
\theta_{\mu\tau}^{\ell}\,\approx\,-\,3\epsilon\,+\,\eta\,
\approx\,\sqrt{m_{\mu}/m_{\tau}}\,\left|\frac{-\,3\epsilon\,+
\,\eta}{3\epsilon\,+\,\eta}\right|^{1/2}\,\simeq\,0.437
 \label{e24} 
\end{eqnarray} 
\vspace*{-2em}
\begin{eqnarray} 
m_{\nu_{e}}^{D}\,\simeq\,[\,9\epsilon^{'2}/(9\epsilon^{2}\,-\,\sigma^{2})]\,m_{U}\,\simeq\,0.4\,\mbox{MeV}
\label{e25} 
\end{eqnarray} 
\vspace*{-2em}
\begin{eqnarray} 
\theta_{e\mu}^\ell\,\simeq\,\left|\frac{\eta'\,-\,3\epsilon'}{\eta'\,+\,3\epsilon'}\right|^{1/2}\,\sqrt{m_{e}/m_{\mu}}\,\simeq\,0.85\,\sqrt{m_{e}/m_{\mu}}\,\simeq\,0.06
\label{e26}
\end{eqnarray} 
\vspace*{-1.5em}
\begin{eqnarray} 
\theta_{e\tau}^\ell\,\simeq\,\frac{1}{0.85}\,\sqrt{m_{e}/m_{\tau}}\,(m_{\mu}/m_{\tau})\,\simeq\,0.0012\,.
\label{e27} 
\end{eqnarray} 
In evaluating $\theta_{e\mu}^\ell$, we have assumed $\epsilon'$ and
$\eta'$ to be relatively positive.

Given the bizarre pattern of quark and lepton masses and mixings, it
seems remarkable that the simple pattern of fermion mass-matrices,
motivated by the group theory of G(224)/SO(10), gives an overall fit
to  all of them which is good to within $10\%$.  This includes the two
successful predictions on $m_{b}$ and $V_{cb}$ (Eqs.(\ref{e18} and
(\ref{e19})). Note that in supersymmetric unified theories, the
``observed'' value of  $m_{b}(m_{b})$ and renormalization-group
studies suggest that, for a wide  range of the parameter $\tan\beta$,
$m_{b}^{0}$ should in fact be  about 10-20$\%$ {\it lower} than
$m_{\tau}^{0}$ \cite{no43a}.  This is  neatly explained by the
relation: $m_{b}^{0}\approx m_{\tau}^{0}(1 - 8\epsilon^{2})$
(Eq. (\ref{e18})), where exact equality holds in the limit
$\epsilon\rightarrow0$ (due to SU(4)-color), while the decrease of
$m^{0}_{b}$ compared to $m^{0}_{\tau}$ by $8\epsilon^{2}\sim10\%$ is
precisely because the off-diagonal $\epsilon$-entry is proportional to
$B$-$L$ (see Eq. (\ref{e16})).    
 
Specially intriguing is the result on $V_{cb}\approx0.045$ which
compares well with the observed value of $\simeq0.04$.  The
suppression  of $V_{cb}$, compared to the value of $0.17 \pm 0.06$
obtained from Eq.  (\ref{e13}), is now possible because the mass
matrices (Eq. (\ref{e16})) contain an antisymmetric component
$\propto\epsilon$. That corrects the square-root  formula
$\theta_{sb}=\sqrt{m_{s}/m_{b}}$ (appropriate for symmetric matrices,
see Eq. (\ref{e11})) by the asymmetry factor
$|(\eta+\epsilon)/(\eta-\epsilon)|^{1/2}$ (see Eq. (19)), and
similarly for the angle $\theta_{ct}$. This factor suppresses $V_{cb}$
if $\eta$ and $\epsilon$ have opposite signs. The interesting point is
that, {\it the same feature necessarily enhances the corresponding
mixing  angle $\theta_{\mu\tau}^{\ell}$ in the leptonic sector}, since
the  asymmetry factor in this case is given by
$[(-3\epsilon+\eta)/(3\epsilon+\eta)]^{1/2}$ (see Eq. (24)). This
enhancement of $\theta_{\mu\tau}^\ell$ helps to account for the nearly
maximal oscillation angle observed at SuperK (as discussed
below). This intriguing correlation between the mixing angles in the
quark versus leptonic sectors -- {\it that is suppression of one
implying enhancement of the other} -- has become possible only because
of the $\epsilon$-contribution, which is simultaneously antisymmetric
and is proportional to $B$-$L$. That in turn becomes possible because
of the group-property of SO(10) or a string-derived G(224)$^{4}$. 

Taking stock, we see an overwhelming set of evidences in favor of
$B$-$L$ and  in fact for the full SU(4)-color-symmetry.    These
include: (i) the suppression of $V_{cb}$, together with the
enhancement of $\theta_{\mu\tau}^{\ell}$, just mentioned above, (ii)
the successful relation  $m_{b}^{0}\approx
m_{\tau}^{0}(1-8\epsilon^{2})$, (iii) the usefulness  again of the
SU(4)-color-relation $m(\nu_{Dirac}^{\tau})^{0}\approx m_{t}^{0}$ in
accounting for $m(\nu_{L}^{\tau})$( see Sec. 4 ),  and (iv) the
agreement of the relation
$|m_{s}^{0}/m_{\mu}^{0}|=|(\epsilon^{2}-\eta^{2})/(9\epsilon^{2}-\eta^{2})|$
with the data, in that the ratio   is naturally {\it less than} 1, if
$\eta\sim\epsilon$.    The presence of  $9\epsilon^2$ in the
denominator is because the off-diagonal entry is  proportional to
B-L. Finally, the need for  ($B$-$L$)- as a local symmetry, to
implement baryogenesis, has been noted in Sec.1.

Turning to neutrino masses, while all the entries in the Dirac mass
matrix $N$ are now fixed, to obtain the parameters for the light
neutrinos, one needs to specify those of the Majorana mass matrix of
the RH neutrinos ($\nu^{e,\mu,\tau}_{R}$). Guided by economy and the
assumption of hierarchy, we consider the following pattern :
\begin{eqnarray} 
M_{\nu}^{R} = \left(\begin{array}{ccc}{x} & {0} & {z} \\ {0} & {0} &
{y} \\ {z} & {y} & {1}\end{array}\right)\,M_{R}\,. 
\label{e28}
\end{eqnarray} 

 As discussed in Sec. 4, the  magnitude of
$M_{R}\approx(5\mbox{-}15)\times10^{14}$ GeV can quite plausibly be
justified in the context of supersymmetric unificaton\footnote{This
estimate for $M_{R}$ is retained even if one allows for
$\nu_{\mu}\mbox{-}\nu_{\tau}$ mixing (see Ref. \cite{no6}).} (e.g. by
using $M\approx M_{st}\approx4\times10^{17}$ GeV in Eq. (\ref{e8})).
 To the same extent,  the
magnitude of $m(\nu_{\tau})\approx(1/10\mbox{-}1/30)$ eV, which is
consistent with the SuperK value, can also be anticipated. Thus there
are effectively three new parameters: $x$, $y$, and $z$. Since there
are six observables for the three light neutrinos,
 one can expect three predictions.
These may be taken to be $\theta_{\nu_{\mu}\nu_{\tau}}^{osc}$,
$m_{\nu_{\tau}}$ (see Eq. (\ref{e10})), and for example
$\theta_{\nu_{e}\nu_{\mu}}^{osc}$.

Assuming successively hierarchical entries as for the Dirac mass matrices,
we presume
that $|y|\sim1/10,
|z|\leq|y|/10$ and $|x|\leq z^{2}$. Now given that $m(\nu_{\tau})\sim1/20$
eV (as estimated in Eq. (\ref{e10})),
the MSW solution for the solar neutrino puzzle \cite{no43b} suggests
that $m(\nu_{\mu})/m(\nu_{\tau})\approx1/10\mbox{-}1/30$. The latter
in turn yields :  $|y|\approx(1/18\mbox{\,\,to\,\,}1/23.6)$, with $y$
having the same sign as $\epsilon$ (see Eq. (\ref{e17})). This
solution for y  obtains only by assuming that $y$ is $O(1/10)$ rather than
$O(1)$. Combining now with the mixing in the $\mu$-$\tau$ sector
determined
above (see Eq. (\ref{e24})), one can then determine the
$\nu_{\mu}\mbox{-}\nu_{\tau}$ oscillation angle. The two predictions
of the model for the neutrino-system are then :
\begin{eqnarray}
m(\nu_{\tau})\,\approx\,(1/10\,\mbox{-}\,1/30)\,\mbox{eV}
\label{e29n}
\end{eqnarray}
\vspace*{-2em}
\begin{eqnarray}
\theta_{\nu_{\mu}\nu_{\tau}}^{osc}\,\simeq\,\theta_{\mu\tau}^{\ell} - 
 \theta_{\mu\tau}^{\nu}\,\simeq
\,\left(0.437\,+\,\sqrt{\frac{m_{\nu_{2}}}{m_{\nu_{3}}}}\,\right)\, .
\label{e29}
\end{eqnarray}
\vspace*{-2em}
\begin{eqnarray}
\mbox{Thus,}\,\,\,{\sin}^{2}\,2\theta_{\nu_{\mu}\nu_{\tau}}^{osc}=
(0.96,0.91,0.86,0.83,0.81)\,\,\,\,\\
\mbox{for}\,\,\,\, m_{\nu_{2}}/m_{\nu_{3}}=(1/10,1/15,1/20,1/25,1/30)\,.
\label{e30}
\end{eqnarray}
Both of these predictions are extremely successful.

Note the interesting point that the MSW solution, together with the
requirement that $|y|$ should have a natural  hierarchical value (as
mentioned above),
lead to $y$ having the same sign as $\epsilon$; that (it turns
out) implies that the two contributions in Eq.(\ref{e29}) must {\it
add} rather than subtract, leading to an {\it almost maximal
oscillation angle\,}\cite{no6}.  The other factor contributing to the
enhancement of $\theta_{\nu_{\mu}\nu_{\tau}}^{osc}$ is, of course,
also the asymmetry-ratio which increases $|\theta_{\mu\tau}^{\ell}|$
from 0.25 to 0.437 (see Eq. (\ref{e24})). We see that  one can derive
rather plausibly a large $\nu_{\mu}\mbox{-}\nu_{\tau}$ oscillation
angle $\sin^{2}\,2\theta_{\nu_{\mu}\nu_{\tau}}^{osc}\geq0.8$, together
with an understanding of  hierarchical masses and mixings of the
quarks and the charged leptons,  while maintaining a  large hierarchy
in the seesaw derived masses
($m_{\nu_{2}}/m_{\nu_{3}}=1/10\mbox{-}1/30$), all within a unified
framework  including both quarks and leptons.  In the example
exhibited  here, the mixing angles for the mass eigenstates of neither
the neutrinos  nor the charged leptons are really large,in that
$\theta_{\mu\tau}^{\ell}\simeq0.437\simeq23^{\circ}$ and
$\theta_{\mu\tau}^{\nu}\simeq(0.18\mbox{-}0.31)\approx(10\mbox{-}18)^{\circ}$,
{\it yet the oscillation angle obtained by combining the two is
near-maximal.}  This contrasts with most previous work, in which a
large  oscillation angle is obtained either entirely from the neutrino
sector  (with nearly  degenerate neutrinos) or almost entirely from
the charged lepton sector.

While $M_{R}\approx(5\mbox{-}15)\times10^{14}$ GeV and $y\approx-1/20$
are better determined, the parameters $x$ and $z$ can not be obtained
reliably at present because very little is known
about observables involving $\nu_{e}$. Taking, for concreteness,
$m_{\nu_{e}}\approx(10^{-5}\mbox{-}10^{-4}$ (1 to few)) eV and
$\theta^{osc}_{e\tau}\approx\theta^{\ell}_{e\tau}-\theta^{\nu}_{e\tau}\approx10^{-3}\pm0.03$
as inputs, we obtain : $z\sim(1$-$5)\times10^{-3}$ and $x\sim($1 to
few)$(10^{-6}\mbox{-}10^{-5})$, in accord with the guidelines of
$|z|\sim|y|/10$ and $|x|\sim z^{2}$. This in turn yields :
$\theta^{osc}_{e\mu}\approx\theta^{\ell}_{e\mu}-\theta^{\nu}_{e\mu}\approx0.06\pm0.015$.
Note that the mass of $m_{\nu_{\mu}}\sim3\times10^{-3}$ eV, that follows
from a
natural hierarchical value for $y\sim-(1/20)$, and $\theta_{e\mu}$ as
above, go well with the small angle MSW explanation\footnote{Although
the small angle MSW solution appears to be more generic within the
approach outlined above, we have found that the large angle solution
can still plausibly emerge in a limited region of parameter space,
without affecting our results on fermion masses.} of the solar
neutrinos puzzle. 

It is worthnoting that although the superheavy Majorana masses of the RH
neutrinos cannot be
observed directly, they can be of cosmological significance.  The
pattern given above and the arguments given in Sec. 3 and in this
section suggests that
$M(\nu_{R}^{\tau})\approx(5\mbox{-}15)\times10^{14}$  GeV,
$M(\nu_{R}^{\mu})\approx(1\mbox{-}4)\times10^{12}$ GeV (for
$x\approx1/20$); and $M(\nu_{R}^{e})\sim(1/2\mbox{-}10)\times10^9$ GeV
(for $x\sim(1/2\mbox{-}10)10^{-6}>z^2$).    A mass of
$\nu_{R}^{e}\sim10^{9}$ GeV is of the   right magnitude for producing
$\nu_{R}^{e}$ following reheating and inducing lepton asymmetry   in
$\nu_{R}^{e}$ decay into $H^{0}+\nu_{L}^{i}$, that is subsequently
converted into baryon asymmetry by the electroweak sphalerons
\cite{no19,no20}.

In summary, we have proposed an economical and predictive pattern for
the Dirac  mass matrices, within the SO(10)/G(224)-framework, which is
remarakbly successful in describing the observed masses and  mixings
of {\it all} the quarks and charged leptons. It leads to five
predictions for just the quark- system, all of which agree with
observation to
within 10\%.  The same  pattern, supplemented with a similar structure
for the Majorana mass matrix, accounts for both the large
$\nu_{\mu}$-$\nu_{\tau}$ oscillation angle and a mass of
$\nu_{\tau}\sim1/20$ eV, suggested by the SuperK data. It also
accomodates a  small $\nu_{e}$-$\nu_{\mu}$ oscillation angle relevant
for  theories of the solar neutrino deficit. Given this degree of
success, it makes good sense to study proton decay concretely within
this SO(10)/G(224)-framework. The results of this study \cite{no6} are
presented in the next section.

Before turning to proton decay, it is worth noting that much of our
discussion  of fermion masses and mixings, including those of the
neutrinos, is  essentially unaltered if we go to the limit
$\epsilon'\rightarrow0$ of  Eq. (28).  This limit clearly involves: 
\begin{eqnarray} 
m_{u}\,=\,0\,,\,\,\,\,\theta_{C}\,\simeq\,\sqrt{m_{d}/m_{s}}\,,\,\,\,\,m_{\nu_{e}}\,=\,0\,,\,\,\,\,\theta_{e\mu}^{\nu}\,=\,\theta_{e\tau}^{\nu}\,=\,0\,.  
\nonumber 
\end{eqnarray}
\begin{eqnarray} 
|V_{ub}|\,\simeq\,\sqrt{\frac{\eta\,-\,\epsilon}{\eta\,+\,\epsilon}}\,\sqrt{m_{d}/m_{b}}\,(m_{s}/m_{b})\,\simeq\,(2.1)(0.039)(0.023)\,\simeq\,0.0019 
\label{e32}
\end{eqnarray} 
All other predictions  remain unaltered.  Now, among the observed
quantities in the list above, $\theta_C\simeq\sqrt{m_{d}/m_{s}}$ is a
good result. Considering that $m_{u}/m_{t}\approx10^{-5}$, $m_{u}=0$
is  also a pretty good result.  There are of course plausible small
corrections  which could arise through Planck scale physics; these
could  induce a small value for $m_{u}$ through the (1,1)-entry
$\delta\approx10^{-5}$.    For considerations of proton decay, it is
worth distinguishing between these two variants, which  we will refer
to as cases I and II respectively. 
\begin{eqnarray} 
\mbox{Case I
:}&&\epsilon'\,\approx\,2\,\times\,10^{-4}\,,\,\,\,\delta\,=\,0
\nonumber \\[1em]
\mbox{Case II :}&&\delta\,\approx\,10^{-5}\,,\,\,\,\epsilon'\,= 0\,. 
\label{e33}
\end{eqnarray}

\section{Expectations for Proton Decay in Supersymmetric Unified Theories} 

{\large{\bf 6.1\,}} Turning to the main purpose of this talk, I
present now the reason why the unification framework based on SUSY
SO(10) or G(224), together with the understanding of fermion masses
and mixings discussed above, strongly suggest that proton decay should
be imminent.

Recall that supersymmetric unfied theories (GUTs) introduce two new
features to proton decay : (i) First, by raising $M_{X}$ to a higher
value of about $2\times10^{16}$ GeV, they strongly suppress the
gauge-boson-mediated $d=6$ proton decay operators, for which
$e^{+}\pi^{0}$ would have been the dominant mode (for this case, one
typically obtains : $\Gamma^{-1}(p\rightarrow
e^{+}\pi^{0})|_{d=6}\approx10^{36\pm1.5}$ yrs). (ii) Second, they
generate $d=5$ proton decay operators \cite{no22} of the form
$Q_{i}Q_{j}Q_{k}Q_{l}/M$ in the superpotential, through the exchange
of color triplet Higginos, which are the GUT partners of the standard
Higgs(ino) doublets, such as those in the $5+\overline{5}$ of SU(5) or
the 10 of SO(10). Assuming that a suitable doublet-triplet splitting
mecahnism provides heavy GUT-scale masses to these color triplets and
light masses to the doublets, these ``standard'' $d=5$ operators,
suppressed by just one power of the heavy mass and the small Yukawa
couplings, are found to provide the dominant mechanism for proton
decay in supersymmetric GUT \cite{no44,no45,no46,no47}.

Now, owing to (a) Bose symmetry of the superfields in $QQQL/M$, (b)
color antisymmetry, and especially (c) the hierarchical Yukawa
couplings of the Higgs doublets, it turns out that these standard
$d=5$ operators lead to dominant $\overline{\nu}K^{+}$ and comparable
$\overline{\nu}\pi^{+}$ modes, but in all cases to highly suppressed
$e^{+}\pi^{0}$, $e^{+}K^{0}$ and even $\mu^{+}K^{0}$ modes. For
instance, for minimal SUSY SU(5), one obtains (with $\tan\beta\leq20$,
say) : 
\begin{eqnarray}
[\,\Gamma(\mu^{+}K^{0})/\Gamma(\overline{\nu}K^{+})\,]^{SU(5)}_{std}\,\sim\,[m_{u}/m_{c}\,\sin^{2}\theta]\,R\,\approx\,10^{-3}\,,
\label{e34}
\end{eqnarray}
where $R\approx0.1$ is the ratio of the relevant $|$matrix
element$|^{2}\times$(phase space), for the two modes.

It was recently pointed out that in SUSY unified theories based on
SO(10) or G(224), which assign heavy Majorana masses to the RH
neutrinos, there exists a new set of color triplets and thereby very
likely a {\it new source} of $d=5$ proton decay operators
\cite{no21}. For instance, in the context of the minimal set of Higgs
multiplets\footnote{The origin of the new $d=5$ operators in the
context of other Higgs multiplets, in particular in the cases where
$126_{H}$ and $\overline{126}_{H}$ are used to break $B$-$L$, has
been discussed in Ref. \cite{no21}.}
$\{45_{H},16_{H},\overline{16}_{H}$ and $10_{H}\}$ (see Sec. 5), these
new $d=5$ operators arise by combining three effective couplings
introduced before :-- i.e., (a) the couplings
$f_{ij}16{i}16{j}\overline{16}_{H}\overline{16}_{H}/M$ (see
Eq. (\ref{e7})) that are required to assign Majorana masses to the RH
neutrinos, (b) the couplings $g_{ij}16{i}16{j}{16}_{H}{16}_{H}/M$,
which are needed to generate non-trivial CKM mixings (see
Eq. (\ref{e15})), and (c) the mass term
$M_{16}16_{H}\overline{16}_{H}$. For the $f_{ij}$ couplings,
there are two possible SO(10)-contractions, and we assume both to have
comparable strength\footnote{One would expect such a
general contraction to hold, especially if the $f_{ij}$ couplings are
induced by non-perturbative quantum gravity. Furthermore, the $f_{ij}$
couplings with the contraction of the pair
($16_{i}\cdot\overline{16}_{H}$), being effectively in 45 (rather than
in 1) of SO(10), would be induced also by tree-level exchanges, if
these pairs couple to the 45's in the string tower. Such a contraction
would lead to proton decay.}. In this case, the
color-triplet Higginos in $\overline{16}_{H}$ and $16_{H}$ of mass
$M_{16}$ can be exchanged between $\tilde{q}_{i}q_{j}$ and
$\tilde{q}_{k}q_{l}$-pairs. This exchange generates a new set of $d=5$
operators in the effective superpotential of the form
\begin{eqnarray}
W_{new}\,\propto\,f_{ij}\,g_{kl}\,(16_{i}\,16_{j})\,
(16_{k}\,16_{l})\,\langle\overline{16}_{H}\rangle\,
\langle{16}_{H}\rangle/M^2 (1/M_{16})\,,
\label{e35}
\end{eqnarray}
which induce proton decay. Note that these operators depend, through
the couplings $f_{ij}$ and $g_{kl}$, both on the Majorana and on the
Dirac masses of the respective fermions. {\it This is why within SUSY
SO(10) or G(224), proton decay gets intimately linked to the masses
and mixings of all fermions, including neutrinos.}

\subsection*{6.2\,\,\,\,Framework for Calculating Proton Decay Rate}

To establish notations, consider the case of minimal SUSY SU(5) and, as 
an example, the process
$\tilde{c}\tilde{d}\rightarrow{\bar{s}}{\bar{\nu}_{\mu}}$, which
induces $p\rightarrow\overline{\nu}_{\mu}K^{+}$. Let the strength of
the corresponding $d=5$ operator, multiplied by the product of the CKM
mixing elements entering into wino-exchange vertices, 
(which in this case is $\sin\theta_{C}\cos\theta_{C})$  
be denoted by $\hat{A}$. Thus (putting $\cos\theta_{C} =1$), one obtains:  
\begin{eqnarray} 
{\hspace{-0.3cm}}\hat{A}_{\tilde{c}\tilde{d}}(SU(5))\,=\,(h_{22}^{u}\,h_{12}^{d}/M_{H_{C}})\,\sin\theta_{c}\,\simeq\,(m_{c}m_{s}\,\sin^{2}\theta_{C}/v_{u}^{2})\,(\tan\beta/M_{H_{C}})\nonumber\\&&{\hspace{-13.3cm}}\simeq\,(1.9\times10^{-8})\,(\tan\beta/M_{H_C})\,\approx\,(2\times10^{-24}\,\mbox{GeV$^{-1}$})\,(\tan\beta/2)\,(2\times10^{16}\,\mbox{GeV}/M_{H_{C}})\,, 
\label{e36} 
\end{eqnarray}
where $\tan\beta\equiv{v}_{u}/v_{d}$, and we have put $v_{u}=174$ GeV
and the fermion masses extrapolated to the unification-scale --
i.e. $m_{c}\simeq300$ MeV and $m_{s}\simeq40$ MeV.  The amplitude for the associated four-fermion process $dus\rightarrow\overline{\nu}_{\mu}$ is given by: 
\begin{eqnarray} 
A_5(dus\,\rightarrow\,\overline{\nu}_{\mu})\,=\,\hat{A}_{\tilde{c}\tilde{d}}\,\times\,(2f)
\label{e37} 
\end{eqnarray} 
where $f$ is the loop-factor associated with wino-dressing. Assuming 
$m_{\tilde{w}}\ll{m}_{\tilde{q}}\sim{m}_{\tilde{l}}$ one gets:
$f\simeq(m_{\tilde{w}}/m^{2}_{\tilde{q}})(\alpha_{2}/4\pi)$.  
Using the amplitude for $(du)(s\nu_\ell$), as in Eq. (\ref{e37}), 
($\ell=\mu$ or $\tau$), one then obtains \cite{no45,no46,no47,no6} :  
\begin{eqnarray} 
\Gamma^{-1}(p\,\rightarrow\,\overline{\nu}_{\tau}K^+)\,\approx\,(2.2\times10^{31})\,\mbox{yrs}\times\nonumber\\&&\hspace{-7cm}\left(\frac{0.67}{{A}_{S}}\right)^2\,\left[\frac{0.006\,\mbox{GeV}^3}{\beta_{H}}\right]^{2}\,\left[\frac{(1/6)}{(m_{\tilde{W}}/m_{\tilde{q}})}\right]^{2}\left[\frac{m_{\tilde{q}}}{1\,\mbox{TeV}}\right]^2\,\left[\frac{2\times10^{-24}\,\mbox{GeV}^{-1}}{\hat{A}(\overline{\nu})}\right]^2\,. 
\label{e38}
\end{eqnarray} 
Here $\beta_{H}$ denotes the hadronic matrix element defined by
$\beta_{H}
u_{L}(\vec{k})\equiv\epsilon_{\alpha\beta\gamma}\langle0|(d_{L}^{\alpha}u_{L}^{\beta})u_{L}^{\gamma}|p,\vec{k}\rangle$.
While the range $\beta_{H}=(0.003\mbox{-}0.03)$ GeV$^{3}$ has been
used in the  past \cite{no46}, given that one  lattice calculation
yields $\beta_{H}=(5.6\pm0.5)\times10^{-3}$ GeV$^3$ \cite{no48},  we
will take as a plausible range : $\beta_{H}=(0.006$
GeV$^3$)$(1/2-2)$.Here,
$A_{S}\approx0.67$ stands for the short distance renormalization
factor of the $d=5$ operator. Note that the 
familiar factors that appear in the expression for proton 
lifetime 
-- i.e., $M_{H_{C}}$, ($1+y_{tc}$) representing the interference between
the 
$\tilde{t}$ and $\tilde{c}$ contributions, and $\tan\beta$ (see
e.g. Ref.\cite{no46}) -- are all 
effectively 
contained in $\hat{A}(\overline{\nu})$. 
Allowing for plausible and rather generous uncertainties in the matrix 
element and the spectrum we take: 
\begin{eqnarray} 
\beta_{H}\,=\,(0.006\,\mbox{GeV}^3)\,(1/2\,\mbox{-}\,2) 
\nonumber 
\end{eqnarray}
\begin{eqnarray} 
m_{\tilde{w}}/m_{\tilde{q}}\,=\,1/6\,(1/2\,\mbox{-}\,2)\,,\,\,\,\,{\rm and}\,\,\,\,m_{\tilde{q}}\,\approx\,m_{\tilde{\ell}}\,\approx\,1\,\mbox{TeV}\,(1/\sqrt{2}\,\mbox{-}\,\sqrt{2})\,. 
\label{e39}
\end{eqnarray}
Using Eqs. (\ref{e38}-\ref{e39}), we get : 
\begin{eqnarray} 
\Gamma^{-1}(p\,\rightarrow\,\overline{\nu}_{\tau}K^{+})\,\approx\,(2.2\times10^{31}\,\mbox{yrs})\,[\,2.2\times10^{-24}\,\mbox{GeV}^{-1}/\hat{A}(\overline{\nu}_\ell)\,]^{2}\,[32\,\mbox{-}\,1/32\,]\,.
\label{e40} 
\end{eqnarray} 
 
This relation, as well as Eq. (\ref{e38}) are general, depending only
on $\hat{A}(\overline{\nu}_{\ell})$ and on the range of parameters
given in  Eq. (\ref{e39}). They can thus be used for both SU(5) and
SO(10).

The experimental lower limit on the inverse rate for the  
$\bar{\nu}K^{+}$ modes is given by \cite{no9},  
\begin{eqnarray} 
[\sum_{\ell}\,\Gamma(p\,\rightarrow\,\overline{\nu}_{\ell}K^{+})]^{-1}_{expt} > 1.6 \times 10^{33}\,\mbox{yrs}\,. 
\label{e41}
\end{eqnarray} 
Allowing for all the uncertainties to stretch in the same  direction
(in this case, the square bracket = 32), and assuming  that just one
neutrino flavor (e.g. $\nu_{\mu}$ for SU(5)) dominates, the   observed
limit Eq. (\ref{e41}) provides an upper bound on the
amplitude\footnote{If there are sub-dominant
$\overline{\nu}_{i}K^{+}$ modes with branching ratio $R$, the right
side of Eq. (\ref{e42}) should be divided by $\sqrt{1+R}$.}:  
\begin{eqnarray} 
\hat{A}(\overline{\nu}_{\ell})\,\leq\,\sqrt{2}\times10^{-24}\,\mbox{GeV}^{-1} 
\label{e42}
\end{eqnarray} 
which holds for both SU(5) and SO(10).  For minimal SU(5), using
Eq. (\ref{e36}) and $\tan\beta\geq2$ (which  is suggested on several
grounds), one obtains a lower limit on $M_{HC}$  given by: 
\begin{eqnarray} 
M_{HC}\,\geq\,3\times10^{16}\,\mbox{GeV}\,\,\,(\mbox{SU}(5)) 
\label{e43}
\end{eqnarray}
At the same time, higher values of $M_{HC}>3\times10^{16}$ GeV do  not
go very well with gauge coupling unification.  Thus keeping
$M_{HC}\leq3\times10^{16}$ and $\tan\beta\geq2$, we  obtain from
Eq. (\ref{e36}):
$\hat{A}(\mbox{SU}(5))\geq(4/3)\times10^{-24}\,\mbox{GeV}^{-1}$.
Using Eq. (\ref{e40}), this in turn implies that 
\begin{eqnarray} 
\Gamma^{-1}(p\,\rightarrow\,\overline{\nu}K^+)\,\leq\,1.5\times10^{33}\,\mbox{yrs}\,\,\,(\mbox{SU}(5))
\label{e44} 
\end{eqnarray} 
This is a conservative upper limit. In practise, it is unlikely  that
all the uncertainties, including that in $M_{HC}$, would stretch in
the same direction to nearly extreme values so as to prolong proton
lifetime.  A more reasonable upper limit, for minimal SU(5), thus
seems to  be:
$\Gamma^{-1}(p\rightarrow\overline{\nu}K^+)(SU(5))\leq(0.7)\times10^{33}$
yrs.  Given the experimental lower limit (Eq. (\ref{e41})), we  see
that minimal SUSY SU(5) is already or almost on the verge of being
excluded by  proton decay-searches.  We have of course noted in Sec. 4
that SUSY  SU(5) does not go well with  neutrino oscillations
observed at  SuperK. 
 
Now, to discuss proton decay in the context of supersymmetric SO(10), it 
is necessary to discuss first the mechanism for doublet-triplet 
splitting.  Details of 
this discussion may be found in Ref. \cite{no6}.  A synopsis is
presented in the appendix.

\subsection*{6.3.\,\,\,\,Proton Decay in Supersymmetric SO(10)}

The calculation of the amplitudes $\hat{A}_{std}$ and $\hat{A}_{new}$
for the standard  and the new operators for the SO(10) model, are
given in detail in Ref.   \cite{no6}.  Here, I will present only the
vresults.  It is found that  the four amplitudes
$\hat{A}_{std}(\overline{\nu}_\tau K^{+})$,
$\hat{A}_{std}(\overline{\nu}_{\mu}K^{+})$,
$\hat{A}_{new}(\overline{\nu}_{\tau}K^{+})$ and
$\hat{A}_{new}(\overline{\nu}_{\mu}K^{+})$ are in fact very comparable
to each other, within about a factor of two, either way. Since there
is no reason to expect a near cancellation between the standard and
the new operators, especially for both $\overline{\nu}_{\tau}K^{+}$
and $\overline{\nu}_{\mu}K^{+}$ modes, we expect the net amplitude
(standard + new) to be in the range exhibited by either one.
Following Ref. \cite{no6}, I therefore present the contributions from
the standard and the new operators separately. Using the upper limit
on $M_{eff}\geq3\times10^{18}$ GeV (see Appendix), we obtain a lower
limit for the standard proton decay amplitude given by
\begin{eqnarray} 
\hat{A}(\overline{\nu}_{\tau}K^{+})_{std}\,\geq\,\left[\begin{array}{c}{(7\times10^{-24}\,\mbox{GeV}^{-1})\,(1/6\,\mbox{-\,1/4)}\,\,\,\,\,\,\,\,\mbox{case
I}}\\{(3\times10^{-24}\,\mbox{GeV}^{-1})\,(1/6\,\mbox{-\,1/2)}\,\,\,\,\,\,\,\,\mbox{case II}}\end{array}\right]
\label{e45} 
\end{eqnarray} 
Substituting into Eq. (\ref{e40}) and adding the contribution from the
second  competing mode $\overline{\nu}_{\mu}K^{+}$, with a typical
branching ratio $R\approx0.3$, we obtain 
\begin{eqnarray} 
\Gamma^{-1}(\overline{\nu}K^{+})_{std}\,\leq\,\left[\begin{array}{c}{(3\times10^{31}\,\mbox{yrs.})\,(1.6\,\mbox{-}\,0.7)}\\{(6.8\times10^{31}\,\mbox{yrs.})\,(4\,\mbox{-}\,0.44)}\end{array}\right]\,(32\,\mbox{-}\,1/32)
\label{e46}
\end{eqnarray} 
The upper and lower entries in Eqs. (\ref{e45}) and (\ref{e46}) 
  correspond to the cases I and II of the fermion
mass-matrix - i.e.  $\epsilon'\neq0$ and $\epsilon'=0$ - respectively,
(see Eq. (\ref{e33})).  The  uncertainty shown inside the square
brackets correspond to that in the  relative phases of the different
contributions.  The uncertainty of (32 to 1/32) arises from that in
$\beta_{H}$,  $(m_{\tilde{W}}/m_{\tilde{q}})$ and $m_{\tilde{q}}$ (see
Eq. (\ref{e39})). Thus we find that for MSSM embedded in SO(10),   the
inverse partial proton decay rate  should satisfy :
\begin{eqnarray} 
\Gamma^{-1}(p\,\rightarrow\overline{\nu}K^{+})_{std}\,\leq\,\left[\begin{array}{c}{3\times10^{31\pm1.7}\,\mbox{yrs.}}\\{6.8\times10^{31^{+2.1}_{-1.5}}\,\mbox yrs.}\end{array}\right]\,\leq\,\left[\begin{array}{c}{1.5\times10^{33}\,\mbox{yrs.}}\\{7\times10^{33}\,\mbox{yrs.}}\end{array}\right]\,\,\,\,(\mbox{SO(10)})\,.
\label{e47} 
\end{eqnarray}
The central value of the upper limit in Eq. (\ref{e47}) corresponds to
taking the upper limit on $M_{eff}$.
The uncertainties of matrix element and spectrum are reflected in the
exponents.The uncertainity in the most sensitive entry of the fermion mass
matrix -
i.e. $\epsilon'$ - is fully incorporated (as regards obtaining an upper
limit on the lifetime) by
going from case I to case II . Note that this increases the lifetime by
almost a
factor of five. Any non-vanishing value of $\epsilon'$
would only shorten the lifetime compared to case II. In this sense,
the larger of the two upper limits quoted above is rather
conservative.
 
Evaluating similarly the contributions from only the new operators, we
obtain
:
\begin{eqnarray}
\Gamma^{-1}(\overline{\nu}K^{+})_{new}\,\approx\,(3\times10^{31}\,\mbox{yrs})\,[16\,\mbox{-}\,1/1.7]\,\{32\,\mbox{-}\,1/32\}\,. 
\label{e48}
\end{eqnarray}
Note that this contribution is independent of $M_{eff}$. It turns out
that it is also insensitive to $\epsilon'$ ; thus it is  nearly the same
for cases I and II. Allowing for a net uncertainty at the upper end by as
much as a factor
of  20
to 200, arising jointly from the square and the curly brackets, i.e.,
without going to extreme ends of all parameters, the new operators
related  to neutrino masses, by  themselves, lead to a proton decay
lifetime bounded by: 
\begin{equation} 
\Gamma^{-1}(\overline{\nu}K^{+})^{expected}_{new}\,\leq\,(0.6\,\mbox{-}\,6)\times10^{33}\,\mbox{yrs.}\,\,\,\mbox{(SO(10)
or string G(224))}
\label{e49} 
\end{equation}
It should be stressed that while the standard $d=5$ operators would be
absent for a string-derived G(224)-model, the new $d=5$ operators,
related to the Majorana masses of the RH neutrinos and the CKM
mixings, would still be present for such a model. {\it Thus our
expectations for the proton decay lifetime (as shown in Eq. (\ref{e49}))
and the prominence of the $\mu^{+}K^{0}$ mode (see below) hold for a
string-derived G(224)-model, just as they do for SO(10).}

\subsection*{6.4.\,\,\,\,The Charged Lepton Decay Mode
$(p\rightarrow\mu^{+}K^{0})$}

I now note a distinguishing feature of the SO(10) or the
G(224) model presented here. 
Allowing for  uncertainties in the way the standard and the
new operators can combine  with each other for the three leading modes
i.e. $\overline{\nu}_{\tau}K^{+}$,  $\overline{\nu}_{\mu}K^{+}$ and
$\mu^{+}K^{0}$, we obtain (see Ref. \cite{no6} for  details): 
\begin{eqnarray} 
B(\mu^{+}K^{0})_{std+new}\,\approx\,\left[1\%\,\,\mbox{to}\,\,50\%\right]\,\rho\,\,\,\,\mbox{(SO(10)
or string G(224))}
\label{e50} 
\end{eqnarray} 
where $\rho$ denotes the ratio of the squares of relevant matrix
elements  for the $\mu^{+}K^{0}$ and $\overline{\nu}K^{+}$ modes. In
the absence of a reliable lattice calculation for the $\bar{\nu}K^{+}$
mode \cite{no48}, one should remain open to the possibility  of
$\rho\approx1/2$ to 1 (say). 
 We find that for a
large range of parameters, the branching ratio $B(\mu^{+}K^{0})$ can
lie in the range of 20 to 40\% (if $\rho \approx1$). This prominence of
the $\mu^{+}K^{0}$ mode for the SO(10)/G(224) model
 is primarily due to contributions from the new
operators. This contrasts sharply with the minimal SU(5) model, in
which the $\mu^{+}K^{0}$ mode is expected to have a branching ratio of
only about $10^{-3}$. In short, prominence of the $\mu^{+}K^{0}$ mode,
if seen, would clearly show the relevance of the new operators, and
thereby reveal the proposed link between neutrino masses and proton
decay \cite{no21}.

\subsection*{6.5.\,\,\,\,Section Summary}

In summary, our study of proton decay has been carried out within the
SO(10) or the G(224)-framework\footnote{As described in Secs. 3 and
5.}, with special attention paid to its dependence on fermion masses
and threshold effects. The study strongly suggests an upperlimit on
proton lifetime, given by
\begin{eqnarray}
\tau_{proton}\,\leq\,(1/2\,\mbox{-}\,1)\times10^{34}\,\,\mbox{yrs}\,,
\label{e51}
\end{eqnarray}
with $\overline{\nu}K^{+}$ being the dominant decay mode. Although
there are uncertainties in the matrix element, in the SUSY-spectrum,
and in certain sensitive elements of the fermion mass matrix,
especially $\epsilon'$ (see Eq. (\ref{e47}) for predictions in cases I
versus II), this upper limit is obtained by allowing for a generous
range in these parameters and stretching all of them in the same
direction so as to extend proton lifetime. In this sense, while the
predicted lifetime spans a wide range, the upper limit quoted above is
quite conservative. In turn, it provides a clear reason to expect that
the discovery of proton decay should be imminent. The implication of
this prediction for a next-generation detector is emphasized in the
next section.

\section{Concluding Remarks}

The preceding sections show that one is now in possession of a set of
facts, which may be viewed as the {\it matching pieces of a puzzle} ; in
that all
of them can be resolved by just one idea - that is grand unification.
 These include : (i) the observed
family-structure, (ii) meeting of the three gauge coulings, (iii)
neutrino oscillations; in particular the mass of $\nu_{\tau}$
(suggested by SuperK), (iv) the intricate pattern of the masses and
mixings of all the fermions, including the smallness of
$V_{bc}$ and the largeness of $\theta^{osc}_{\nu_{\mu}\nu_{\tau}}$,
and (v) the need for $B$-$L$ to implement
baryogenesis. All these pieces fit beautifully together within a
single puzzle board framed by supersymmetric unification, based on 
 SO(10) or a
string-unified G(224)-symmetry. 

The one and the most notable piece of the puzzle still missing, 
 however, is proton
decay. Based on a systematic study of this process within the
supersymmetric 
SO(10)/G(224)-framework \cite{no6}, which is clearly favored by the 
data, I have argued here that a conservative upper limit on the proton
lifetime is about (1/2 - 1)$\times10^{34}$ yrs. So,
unless the fitting of all the pieces listed above is a mere coincidence,
and I
believe that that is highly unlikely, discovery of proton decay should
be around the corner. In particular, as mentioned in the Introduction,
we expect that candidate events should be observed in the near future
already at SuperK. However, allowing for the possibility that proton
lifetime may well be near the upper limit stated above, a
next-generation detector providing a net gain in sensitivity by a
factor five to ten, compared to SuperK, would be needed to produce real
events and distinguish them unambiguously from the background. Such an
improved detector would of course be essential to study the branching
ratios of certain crucial though sub-dominant decay modes such as the
$\mu^{+}K^{0}$ .

The reason for pleading for such improved searches is that proton
decay would provide us with a wealth of knowledge about physics at
truly short distances ($<10^{-30}$ cm), which cannot be gained by any
other means. Specifically, the observation of proton decay, at a rate
suggested above, with $\overline{\nu}K^{+}$ mode being dominant, would
not only reveal the underlying unity of quarks and leptons  but also
the relevance of supersymmetry. It would also confirm a unification of
the fundamental forces at a scale of order $2\times10^{16}$ GeV.
Furthermore, prominence of the $\mu^{+}K^{0}$ mode, if seen, would
have even deeper significance, in that in addition to supporting  the
three features mentioned above, it would also reveal the  link between
neutrino masses and proton decay, as discussed in Sec. 6.  {\it In
this sense, the role of  proton decay in probing
into physics
at the
most fundamental level is unique }. In view of  how valuable
such a probe would be and the fact that the predicted  upper limit on
the  proton lifetime is only a factor of three to six higher than the
empirical lower limit, the argument in favor of building an improved
detector seems compelling.

To conclude,
the discovery of proton decay would undoubtedly
constitute a landmark in the history of physics. It would provide the
last, missing piece of gauge unification and would shed light on how such
a unification may be extended to include gravity.

\vskip1.5em

{\bf Acknowledgements :} I would like to thank Kaladi S. Babu and
Frank Wilczek for a most enjoyable collaboration, and Joseph Sucher
for valuable discussions. I would also like to thank the organizers of
the NNN99 workshop, especially Chang Kee Jung, Millind Diwan and Hank
Sobel, for arranging a stimulating meeting and also for the kind
hospitality. The research presented here is supported in part by DOE
grant no. DE-FG02-96ER-41015.

\appendix 
 
\section*{Appendix\\A Natural Doublet-Triplet Splitting Mechanism in
SO(10)} 
 
\setcounter{equation}{0} 
 
\renewcommand{\theequation}{A\arabic{equation}}

In supersymmetric SO(10), a  natural doublet--triplet splitting can be
achieved by coupling the adjoint  Higgs ${\bf 45_{H}}$ to a ${\bf
10_{H}}$ and a ${\bf 10'_{H}}$, with  ${\bf 45_{H}}$ acquiring a
unification--scale VEV in the $B$-$L$ direction  \cite{no49}:
$\langle{\bf 45_{H}}\rangle=(a,a,a,0,0)\times\tau_{2}$ with
$a\sim{M}_{U}$.  As discussed in Section 2, to generate  CKM mixing
for fermions we require  $({\bf 16_{H}})_d$ to acquire a VEV of the
 electroweak scale. To ensure accurate gauge coupling unification,
the effective low energy theory should not contain split multiplets
beyond those of MSSM. Thus the MSSM Higgs doublets must  be linear
combinations of the SU(2$)_{L}$ doublets in ${\bf 10_{H}}$  and ${\bf
16_{H}}$. A simple set of superpotential terms that ensures this  and
incorporates doublet-triplet splitting is \cite{no6}: 
\begin{eqnarray} 
W_{H}\,=\,\lambda\,{\bf 10_{H}\,45_{H}\,10'_{H}}\,+\,M_{10}\,{\bf
10'_{H}}^{2}\,+\,\lambda'\,\overline{\bf 16}_{H}\,\overline{\bf
16}_{H}\,{\bf 10}_{H}\,+\,M_{16}\,{\bf 16}_{H}\overline{\bf 16}_{H}\,.
\label{a1} 
\end{eqnarray}
A complete superpotential for ${\bf 45_{H}}$, ${\bf 16_{H}}$,  ${\bf
\overline{16}_{H}}$, ${\bf 10}_{H}$, ${\bf 10}'_{H}$ and possibly
other fields,  which  ensure that ${\bf 45_{H}}$, ${\bf 16_{H}}$  and
${\bf \overline{16}_{H}}$ acquire unification  scale VEVs with
$\langle{\bf 45_{H}}\rangle$ being  along the $(B$-$L)$ direction,
that exactly two Higgs doublets  $(H_{u},H_{d})$ remain light, with
$H_{d}$ being a linear combination  of $({\bf 10_{H}})_{d}$ and $({\bf
16_{H}})_{d}$, and that there are  no unwanted pseudoGoldstone bosons,
can be constructed. With $\langle{\bf 45_{H}}\rangle$ in the $B$-$L$
direction, it  does not contribute to the Higgs doublet mass matrix, so
one pair
of  Higgs doublet remains light, while all triplets acquire
unification  scale masses.  The light MSSM Higgs doublets are 
\begin{eqnarray} 
H_{u}\,=\,{\bf 10}_{u}\,,\,\,\,\,H_{d}\,=\,\cos\gamma\,{\bf
10}_{d}\,+\,\sin\gamma\,{\bf 16}_{d}\,,
\label{a2} 
\end{eqnarray}
with
$\tan\gamma\equiv\lambda'\langle{\bf\overline{16}_{H}}\rangle/M_{16}$.
Consequently, $\langle{\bf10}\rangle_{d}=(\cos\gamma)\,v_{d}$,
$\langle{\bf16}_{d}\rangle=(\sin\gamma)\,v_{d}$, with $\langle
H_{d}\rangle=v_{d}$ and $\langle{\bf 16}_{d}\rangle$ and $\langle{\bf
10}_{d}\rangle$ denoting the electroweak VEVs of those
multiplets. Note that $H_{u}$ is purely in ${\bf 10_H}$ and that
$\left \langle {\bf 10}_d \right \rangle^2 + \left \langle {\bf 16}_d
\right \rangle^2 = v_d^2$. This mechanism of doublet-triplet (DT)
splitting is rather unique for the minimal Higgs systems in that it
meets the requirements of both D-T splitting and CKM-mixing.  In turn,
it has three special consequences:

(i) It modifies the familiar SO(10)-relation
$\tan \beta \equiv v_u/v_d = m_t/m_b \approx 60$ to:
\begin{eqnarray}
\tan \beta/\cos \gamma \approx m_t/m_b \approx 60
\end{eqnarray}
As a result, even
low to moderate values of $\tan \beta
\approx 3$ to 10 (say) are perfectly allowed in SO(10) (corresponding to
$\cos \gamma \approx 1/20$ to $1/6$).

(ii) The most important consequence of the DT-splitting mechanism
outlined above is this: In contrast to SU(5), for which the strengths
of the standard d=5 operators are proportional to $(M_{H_c})^{-1}$
(where $M_{H_C}\sim few \times 10^{16}$ GeV (see Eq. (\ref{e43})), for
the SO(10)-model, they become proportional to $M_{eff}^{-1}$, where
$M_{eff} =(\lambda a)^2/M_{10'} \sim M_U^2/M_{10'}$. $M_{10'}$ can be
naturally smaller (due to flavor symmetries) than $M_U$ and thus
$M_{eff}$ correspondingly larger than $M_{U}$ by one to two orders of
magnitude (see Ref. \cite{no6}). Now the proton decay amplitudes for
SO(10) in fact possess an intrinsic enhancement compared to those for
SU(5), owing primarily due to differences in their Yukawa couplings
for the up sector (see Appendix C in Ref. \cite{no6}). As a result,
these larger values of $M_{eff}\sim10^{18}$ GeV are in fact needed
for the SO(10)-model to be compatible with the observed limit on the
proton lifetime. At the same time, being bounded above (see below),
they allow optimism as regards future observation of proton decay.

(iii) $M_{eff}$ gets bounded above by considerations of coupling
unification and GUT-scale threshold effects. Owing to mixing
between $10_d$ and $16_d$ (see Eq. (\ref{a2})), the threshold
correction to $\alpha_{3}(m_{z})$ due to doublet-triplet splitting
becomes proportional to $\ln\,(M_{eff}\cos\gamma/M_{U})$. Inclusion of
this correction and those due to  splittings within the gauge and the
Higgs multiplets (i.e. $45_{H}$, $16_{H}$, and
$\overline{16}_{H}$)\footnote{The correction to $\alpha_{3}(m_z)$ due
to Planck scale  physics through the effective operator
$F_{\mu\nu}F^{\nu\mu}45_{H}/M$  vanishes due to antisymmetry in the
SO(10)-contraction.}, together with  the observed degree of coupling
unification allows us to obtain a  conservative upper limit on
$M_{eff}$, given by \cite{no6} :
\begin{eqnarray}
M_{eff}\,\leq\,3\times10^{18}\,\mbox{GeV}\,.
\label{a3} 
\end{eqnarray}  
This in turn helps provide an upper limit on the expected proton decay
lifetime (see text).

\end{document}